\documentclass[twocolumn,tradiabstract]{aa} 

\usepackage{natbib}
\usepackage{graphicx}
\usepackage{amsmath}

\newcommand{\pd}[2]{\frac{\partial #1}{\partial #2}}

\newcommand\mancha{\textsc{Mancha3D~}}
\newcommand\batt{\textsc{batt}}
\newcommand\ambi{\textsc{ambi}}
\newcommand\ambihall{\textsc{ambihall}}

\begin{document}

\title{Joint action of Hall and ambipolar effects in 3D magneto-convection simulations of the quiet Sun. I. Dissipation and generation of waves}
\titlerunning{Magneto-convection with Hall and ambipolar effects}

\author{P. A. Gonz\'alez-Morales\inst{1, 2}, E. Khomenko\inst{1,2}, N. Vitas\inst{1,2},  M. Collados\inst{1,2}}
\authorrunning{Gonz\'alez-Morales et al.}

\institute{Instituto de Astrof\'{\i}sica de Canarias, 38205 La Laguna, Tenerife, Spain
\and Departamento de Astrof\'{\i}sica, Universidad de La Laguna, 38205, La Laguna, Tenerife, Spain}

\date{Received; Accepted }

\abstract {The partial ionization of the solar plasma causes several nonideal effects such as the ambipolar diffusion, the Hall effect, and the Biermann battery effect. Here we report on the first three-dimensional realistic simulations of solar local dynamo where all three effects were taken into account. The simulations started with a snapshot of already saturated battery-seeded dynamo, where two new series were developed: one with solely ambipolar diffusion and another one also taking into account the Hall term in the generalized Ohm's law. The simulations were then run for about 4 hours of solar time to reach the stationary regime and improve the statistics. In parallel, a purely MHD dynamo simulation was also run for the same amount of time. The simulations are compared in a statistical way.  We consider the average properties of simulation dynamics, the generation and dissipation of compressible and  incompressible waves, and the magnetic Poynting flux. The results show that, with the inclusion of the ambipolar diffusion, the amplitudes of the incompressible perturbations related to Alfv\'en waves are reduced, and the Poynting flux is absorbed, with a frequency dependence. The Hall effect causes the opposite action:nsignificant excess of incompressible perturbations is generated and an excess of the Poynting flux is observed in the chromospheric layers.
The model with ambipolar diffusion shows, on average, sharper current sheets and slightly more abundant fast magneto-acoustic shocks in the chromosphere. The model with the Hall effect has higher temperatures at the lower chromosphere and stronger and more vertical magnetic field concentrations all over the chromosphere. The study of high-frequency waves reveals that significant power of incompressible perturbations is associated with areas with intense and more vertical magnetic fields and larger temperatures. This behavior explains the large Poynting fluxes in the simulations with the Hall effect and provides confirmation as to the role of Alfv\'en waves in chromospheric heating in internetwork regions, under the action of both Hall and ambipolar effects. We find a positive correlation between the magnitude of the ambipolar heating and the temperature increase at the same location after a characteristic time of $10^2$ sec.}

\keywords{Sun: photosphere -- Sun: chromosphere --Sun: magnetic field -- Sun: numerical simulations}

\maketitle

\section{Introduction} %

Understanding the magnetic connectivity through the extremely weakly ionized solar photosphere and partially ionized chromosphere is one of the challenging questions in solar physics. A large amount of energy is transported through the solar convection zone. Reaching the surface, this energy is partly transmitted upwards in the form of waves along magnetic fields connecting with the upper atmospheric layers. A nonideal plasma behavior at the surface significantly affects this transport \citep[see the recent review by][]{Ballester+etal2018}. 

The main nonideal mechanisms in the Sun are the ambipolar diffusion, the Hall effect, and the Biermann battery effect, as expressed by the generalized Ohm's law \citep{Khomenko+etal2014b}. The ambipolar diffusion is the only one of them which is, strictly speaking, related to the presence of neutrals. 
The ambipolar diffusion effect is produced by the net relative motion between the neutral and charged populations,  where neutrals are able to move across the magnetic field. Even if the charges are tied to the magnetic field, the center-of-mass of the fluid can move across the magnetic field \citep{Spitzer1962}. The ambipolar diffusion has been studied most extensively in the solar context, both using the analytical theory or idealized, and even realistic, numerical simulations \citep[see e.g.,][among many others]{Piddington1956, Osterbrock1961, DePontieu1998,  Pandey2008, Soler2009,  Soler2015,  Song+Vasyliunas2011, Diaz+etal2013, Khomenko+etal2014, Leake+Arber2006, Arber2007, MartinezSykora+etal2012, MartinezSykora2016, MartinezSykora2017, Khomenko+Collados2012, Shelyag+etal2016, Khomenko+etal2018}. It has been shown that ambipolar diffusion allows one to damp incompressible magnetic waves and currents as well as to convert magnetic perturbations into plasma thermal energy \citep[this additional energy can be spent both in the ionization of the plasma and in the actual increase of the plasma temperature, see ][]{Khomenko+etal2018}. Ambipolar diffusion also modifies the growth rates of instabilities \citep{Soler+etal2012, Diaz+etal2013, Khomenko+etal2014, Ballai+etal2017, Ruderman+etal2018}, and it affects the amount of the emerged flux through the surface \citep{Leake+Arber2006, Arber2007}.  

Recent 2D simulations of flux emergence using realistic physics show that ambipolar heating allows to remove the cold bubbles produced by the adiabatic expansion of chromospheric material \citep{MartinezSykora2017}. However, since ionization rates are much faster than the recombination rates, the nonequilibrium ionization may result in an insufficient action of the ambipolar diffusion, leading to lower heating rates \citep{Nobrega-Siverio+etal2020}. On the other hand, realistic 3D simulations of solar dynamo and magneto-convection of quiet solar regions suggest that heating through ambipolar diffusion might play an important role in upper photospheric layers, where the plasma is denser and the energy is spent directly into a temperature increase and not into ionization \citep[][we refer to this work as Paper I]{Khomenko+etal2018}. This is achieved by the highly intermittent, mixed polarity, and tangled magnetic fields existing everywhere in the quiet Sun and creating a reservoir of magnetic energy that can be dissipated and contribute to the chromospheric energy budget. Nevertheless, only few realistic simulations including ambipolar diffusion exist so far, and only for few scenarios (quiet Sun, flux emergence), and a significant effort is still needed to understand the physics of interaction of neutrals and the solar plasma in the photosphere and chromosphere. 

In the upper chromosphere, collisions may not be strong enough and the difference in the velocities of ions and neutrals may reach some fraction of the sound speed. In this situation, a simplified single-fluid treatment may not be sufficient and multi-fluid modeling should be applied. In a recent study, \citet{Popescu+etal2019a, Popescu+etal2019b} have shown how the ion-neutral decoupling visibly affects waves in neutrals and charges in the chromosphere, and how these effects become more pronounced in shocks. Shocks are present everywhere in the chromosphere, at any moment of time, and the two-fluid effects must be incorporated into the future modeling \citep[as in e.g.,][]{Hillier+etal2016, Snow+Hillier2019}.  Other processes, such as instabilities and reconnection might also acquire a two fluid nature \citep{Leake+etal2012, Leake+etal2014, Murphy+Lukin2015, Alvarez-Laguna+etal2017, Hillier2019}. These processes have only been studied using idealized experiments with a limited application to the chromosphere. Realistic multi-fluid modeling still remains a challenge due to its complexity \citep[see recent works by ][]{Maneva+etal2017, Kuzma+etal2019}.

Yet, another effect that has been studied very little using realistic simulations applied to the Sun is the Hall effect. The Hall effect by itself is not related to the presence of neutrals and exists in any plasma. It acts on scales similar to ion-cyclotron scales and is generally considered not important for the large scale processes in the solar atmosphere that happen on temporal scales of mHz and spatial scales of km or Mm. However, in a partially ionized plasma the Hall effect is amplified by the presence of neutrals. The importance of this effect is measured in units of the so-called Hall parameter $\epsilon$, 
\begin{equation} \label{eq:epsilon1}
\epsilon=\frac{\omega}{\omega_{ci}\xi_i}, 
\end{equation}
which is the ratio between the wave frequency, $\omega$, and the ion-cyclotron frequency $\omega_{ci}$, scaled with the ionization fraction of the medium, $\xi_i=\rho_i/\rho$ \citep{Cheung+Cameron2012, Cally+Khomenko2015}. Since the ionization fraction in the solar photosphere and the chromosphere is typically very low, the range of frequencies for which the effect is important is translated to considerably lower values \citep{Pandey+Wardle2008, Pandey2008, Cheung+Cameron2012}. Weak magnetic field strengths also increases $\epsilon$ through the dependence of $\omega_{ci}$ on B.

Another interpretation of the Hall parameter can be done in terms of the Hall coefficient, 
\begin{equation} \label{eq:eta-hall}
\eta_H=\frac{|B|}{en_e}, 
\end{equation}
which appears in the generalized induction equation in front of the Hall term, see Eq. \ref{eq:induction} later on in Section \ref{sect:model}. The typical time scale of the Hall effect can be defined through $\eta_H$, and this scale can be related to $\epsilon$, as
\begin{equation} \label{eq:epsilon2}
t_{\rm Hall}=\mu_0\frac{\eta_H}{v_A^2}=\frac{\epsilon}{\omega}, 
\end{equation}
where $v_A$ is the single-fluid Alfv\'en speed (that is, using the total density, taking into account all species, neutrals and charges). This way it becomes clear how $\epsilon$, which determines the efficiency of the Hall effect, is related to the Hall coefficient $\eta_H$.  It is also to be noted that the time scale where the Hall effect becomes efficient is higher where the Alfv\'en speed is lower.

In a recent study, \citet{Cally+Khomenko2015} proposed a new mechanism of generation of Alfv\'en waves in the solar photosphere through the action of the Hall term in a partially ionized plasma. The Hall term creates currents, in a natural way, that are transverse to the propagation direction of the waves, opening the possibility for the coupling between the fast magneto-acoustic and Alfv\'en modes in a low-$\beta$ plasma. The wave modes become coupled through the so-called Hall region, where the Hall parameter $\epsilon$ is large, and there is a precession between nearly degenerate fast magneto-acoustic and Alfv\'en modes, and the final state that emerges depends on the parameters of the Hall region. Further numerical simulations of this effect, in an application to the solar atmosphere, done by \citet{Gonzalez-Morales+etal2019}, have shown that this mechanism is indeed able to produce Alfv\'en waves through the mode transformation. In their work, the authors considered the complete chain of wave transformations, from fast (essentially acoustic) mode in the high-$\beta$ photospheric layers into the fast (mainly magnetic) and slow (mainly acoustic) modes, and further to the Alfv\'en mode in the low-$\beta$ chromospheric layers. The efficiency of this process was shown to depend on the relative location between the layer where the plasma $\beta=1$\footnote{The definition of the plasma $\beta$ in wave studies, as the ratio between the squared sound and Alfv\'en speeds, is different from the classical definition of the ratio between the gas to magnetic pressure. However both definitions are only a factor $\gamma/2$ different, very close to one. Since in the Sun the $\beta \approx 1$ region is a time-dependent corrugated surface with a width much smaller than the typical wavelengths, for all practical purposes both definitions are almost indistinguishable. }, and the height where the Hall parameter $\epsilon$ is maximum. If the maximum of $\epsilon$ is reached in the low-$\beta$ atmosphere, Alfv\'en waves are generated efficiently at chromospheric heights and can be rather important for 0.1--1 Hz waves, as compared to the chromospheric energy losses. Obviously, the model considered had simplified assumptions, like a homogeneous inclined field and a fixed time-independent thermodynamic structure of the atmosphere. It is nevertheless interesting because it shows how ion-neutral effects can help produce upward flux of Alfv\'en waves to the upper layers, this way participating in the chain of the energy transport to the chromosphere and corona.

Only few 2D realistic magneto-convection simulations incorporating the  Hall effect have been reported to far \citep{MartinezSykora+etal2012, Cheung+Cameron2012}. It was shown that weak, and out of the plane, currents are created by the Hall term \citep{Cheung+Cameron2012}, in a similar way as described above. Nevertheless, the full action of the Hall effect can only be accessed in 3D. Here we report on such simulations. We continue the series of modeling reported in \citet{Khomenko+etal2018} (Paper I) with the aim to make progress on the investigation of the ion-neutral effects on the energy balance of the solar chromosphere. Here we have performed 3D simulations of battery-excited dynamo including both, ambipolar diffusion and the Hall effects, using the generalized Ohm's law. We statistically compare pairs of simulation runs with and without ambipolar diffusion and with/without the Hall effect. The simulation runs are performed for sufficiently long time to make sure the regime is stationary and the results are statistically valid. Results regarding the Poynting flux absorption as a function of frequency and magnetic structure are presented. We also study the generation and dissipation of different types of waves and their relation to the magnetic field structure. The effects on the average chromospheric temperature increase, and on the magnetic field energy reaching higher atmosphere are discussed.

\section{Battery-excited dynamo simulations}
\label{sect:model}

The simulations studied here represent a continuation of the battery-excited local solar dynamo series from \citet{Khomenko+etal2017} and \citet{Khomenko+etal2018}. The simulations are done with the \mancha code \citep{Khomenko+Collados2006, Felipe+etal2010, Gonzalez-Morales+etal2018} and are fully described in the publications mentioned above. Compared to Paper I, the models include the Hall term, apart from the already present battery and ambipolar terms. The following system of equations is solved:

\begin{eqnarray}
\label{eq:continuity} 
\pd{\rho_1}{t} + \mathbf{\nabla} \cdot \left( \rho\mathbf{v} \right) =  \left( \pd{\rho_1}{t}\right)_{\rm diff},
\end{eqnarray}

\begin{eqnarray} \label{eq:momentum} 
\pd{\rho\mathbf{v}}{t} &+& \mathbf{\nabla}\cdot \left[\rho\mathbf{v} \mathbf{v} + \left(p_1+ \frac{\mathbf{B}_1^2 + 2\mathbf{B}_1 \cdot \mathbf{B}_0}{2 \mu_0} \right) \mathbf{I} \right] + \nonumber \\
&+&\mathbf{\nabla}\cdot \left[\frac{1}{\mu_0} \left(\mathbf{B}_0 \mathbf{B}_1-\mathbf{B}_1 \mathbf{B}_0-\mathbf{B}_1 \mathbf{B}_1 \right) \right] \nonumber \\
&=& \rho_1 \mathbf{g} + \left( \pd{\rho\mathbf{v}}{t}\right)_{\rm diff} ,
\end{eqnarray}

\begin{eqnarray} \label{eq:induction}
\pd{\mathbf{B}_1}{t}  &=&  \mathbf{\nabla}\times \left[\mathbf{v}\times \mathbf{B} \right] + \nonumber \\ &+&  \mathbf{\nabla}\times \left[\frac{\mathbf{\nabla}p_e}{e n_e} -\eta_H\frac{(\mathbf{J} \times \mathbf{B})}{|B|} + \eta_A\frac{\left(\mathbf{J} \times \mathbf{B} \right) \times \mathbf{B} }{|B|^2}   \right]  \nonumber \\
&+& \left(\pd{B_1}{t}\right)_{\rm diff},
\end{eqnarray}

\begin{eqnarray} \label{eq:energy}
\pd{e_1}{t} &+& \nabla \cdot \left[ \mathbf{v}\left(e + p + \frac{|\mathbf{B}|^2}{2 \mu_0} \right) - \frac{\mathbf{B}(\mathbf{v} \cdot \mathbf{B}) }{\mu_0} \right]   =  \nonumber \\
&=& \rho \left(\mathbf{g} \cdot \mathbf{v}\right)  +\mathbf{\nabla}\cdot \left[ \eta_A \frac{\mathbf{B} \times \mathbf{J_\perp} }{\mu_0}   + \frac{\mathbf{\nabla}p_e \times \mathbf{B}} {e n_e \mu_0}    \right] \nonumber \\
&+& Q_{\rm R} + \left( \pd{e_1}{t}\right)_{\rm diff} . 
\end{eqnarray}
These equations are written for perturbations, as solved by the code. The variables have the following meaning: $\rho=\rho_0+\rho_1$ for the mass density, $p=p_0+p_1$ for pressure, $\mathbf{B}=\mathbf{B}_0+\mathbf{B}_1$ for the magnetic field vector, $\mathbf{v}=\mathbf{v}_1$ for the velocity vector, and  $e=e_0+e_1=(\rho_0+\rho_1)\mathbf{v}^2/2 + (\mathbf{B}_0+\mathbf{B}_1)^2/2/\mu_0 + (e_{\rm int 0}+e_{\rm int 1})$ for the total energy. The variables with subscript ``1'' are considered perturbations, and those with the subscript ``0'' are the background values. The latter must fulfill the requirements of a (M)HS equilibrium. The choice of this equilibrium is described in Paper I. The terms labeled  ``diff'' are numerical hyper-diffusion terms, calculated as in \citet{Vogler2005, Felipe+etal2010}.

Equations (\ref{eq:induction}) and (\ref{eq:energy}) contain the ambipolar diffusion terms with the coefficient, $\eta_A$, calculated as
\begin{equation} \label{eq:etaa}
\eta_A =\frac{\xi_n^2|B|^2}{\alpha_n},
\end{equation}
in units of $[ml^3/tq^2]$. Here $\xi_n=\rho_n/\rho$ is the neutral fraction, and $\alpha_n$ is the collisional parameter. The collisional frequencies entering $\alpha_n$ are computed following standard expressions from \citet{Braginskii1965, Draine1986, Lifschitz, Rozhansky}, see also the review by \citet{Ballester+etal2018} and Paper I.

The Hall coefficient, $\eta_H$, in the same units as $\eta_A$, is defined through Eq. (\ref{eq:eta-hall}). We note that despite the units of $\eta_H$, the Hall effect is not related to diffusivity and, in consequence,  does not have the corresponding counterpart in the energy equation, Eq. (\ref{eq:energy}).

The system of equations is closed with an equation of state (EOS) for the solar chemical mixture given by \cite{1989GeCoA..53..197A}, as described in Paper I. The electron and neutral fraction is stored in the same lookup tables as the rest of the variables needed for the EOS, and is updated self-consistently at every computational step, assuming instantaneous ionization. 

The radiative loss term $Q_{\rm R}$ is computed by solving the radiative transfer (RT) equation, assuming local thermodynamic equilibrium (LTE). The numerical solution is done using short characteristics. The angle integration is done using quadrature with three rays per octant. Here we use the so-called ``gray'' approximation, that is, wavelength-independent opacity. The details of the $Q_{\rm R}$  calculations in \mancha code are given in Paper I.

\begin{figure*}
\centering
\includegraphics[keepaspectratio,width=16cm]{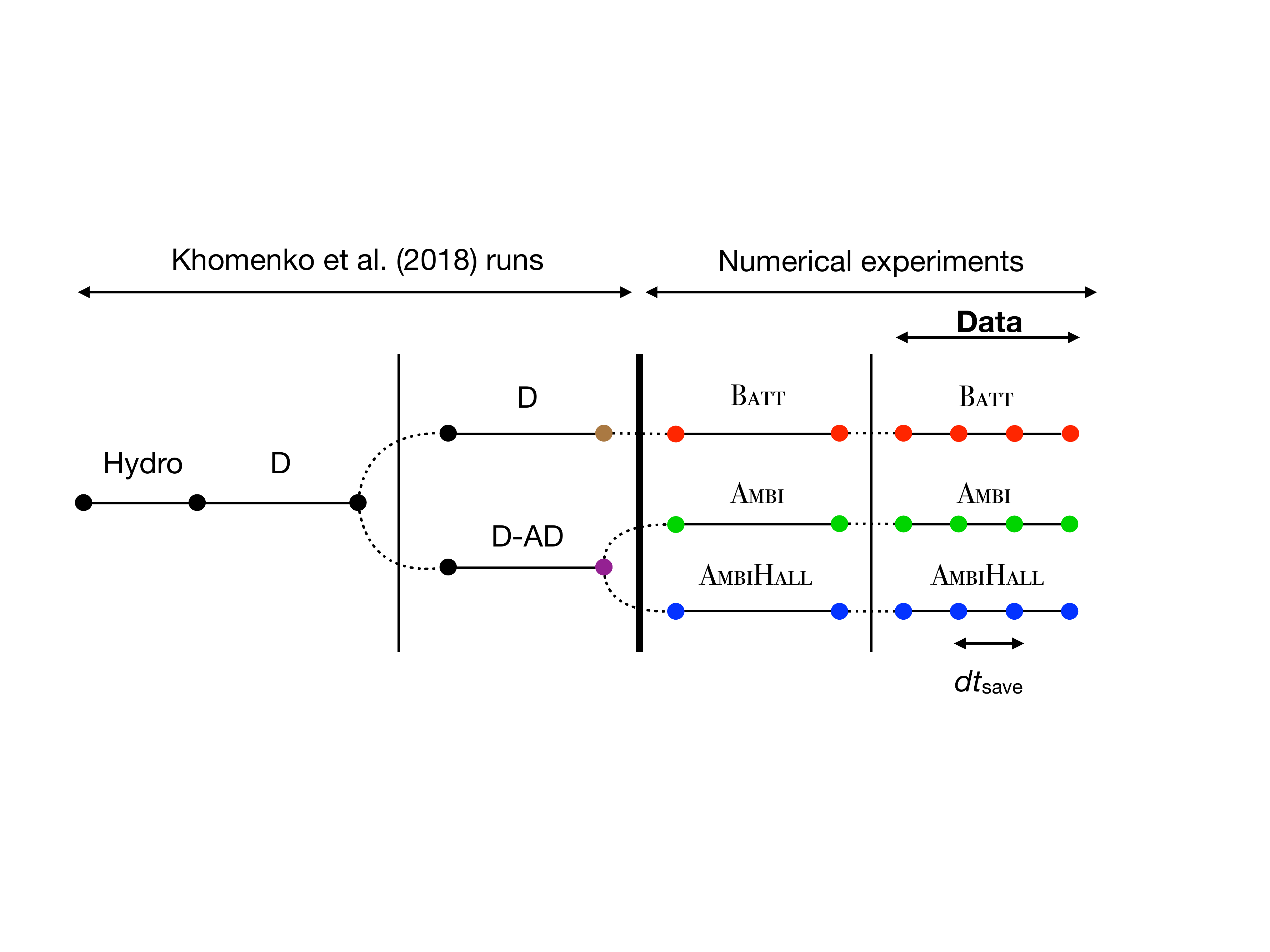}
\caption{\footnotesize Schematic evolution of the numerical experiments and the origin of different branches. The \batt\ simulation has its origin in the D run of \cite{Khomenko+etal2018}. Meanwhile, the simulations \ambi\ and \ambihall\ have their origin in the simulation D-AD of \citet{Khomenko+etal2018}. The analyzed data in this work correspond to the last 2h of each branch using a saving cadence of 10 s.}\label{fig:NumExpData}
\end{figure*}

\subsection{Numerical setup}\label{sect:setup}

We analyze three different runs here, that are referred to as \batt, \ambi\ and \ambihall. All of them are initiated as purely hydrodynamical convection. We use as a background atmosphere (parameters with the subscript ``0'' in the Equations (\ref{eq:continuity})--(\ref{eq:energy})) a plane-parallel model of the Sun's interior and the upper layers, combining the model of  \citet{Spruit1974} at heights [-0.95, 0] Mm below the surface with the HSRA model \citep{hsra} at heights [0, 1.4] Mm above the surface. The horizontal extent of the simulation box is  5.8$\times$5.8 Mm. The simulation box is covered by a uniform grid of 20 km in both horizontal directions, and 14 km in the vertical direction. 

The upper boundary is closed to mass flows, having zero gradient in density and internal energy; the temperature is computed using the tabulated EOS and, because it has no further restriction imposed, the atmosphere is structured by shock viscous dissipation, magnetic field dissipation, and radiative losses. The bottom boundary is open to mass flows, keeping an approximately constant value for the mass inside the domain by controlling its global fluctuations and the total radiative output with zero magnetic field inflow. The rest of the technical details about the simulations, including the boundary conditions and the mass and energy controls, are fully described in \citet{Khomenko+etal2017} and \citet{Khomenko+etal2018}.

The background model is perturbed by random noise to initiate the convection, as described in \citet{Khomenko+etal2017} and \citet{Khomenko+etal2018}. When the convection reaches the stationary regime, after about 3.4 hours of solar time, the battery term is introduced in Eqs. (\ref{eq:induction}) and (\ref{eq:energy}). This term ``seeds'' an initial magnetic field with a strength of about 10$^{-6}$ G.  Its continuous action, together with dynamo amplification, provides the magnetization of the model with  $\sim10^2$ G mean field strength at the $\tau_5$=1 surface, after reaching the saturated regime in about 2 hours of solar time \citep[see Figure 3 in][]{Khomenko+etal2017}. The model is run for 4.2 hours of solar time after switching on the battery term, with the last 2 hours being in the stationary saturated dynamo regime.

At this point we take the snapshot of the dynamo run and switch on the ambipolar term in Eqs. (\ref{eq:induction}) and (\ref{eq:energy}). The simulation is then run for another 150 min of solar time. The last 50 min of this series, with snapshots saved every 20 sec, were analyzed in \citet{Khomenko+etal2018}. This run is labeled as D-AD.  In parallel, the dynamo simulation is run under exactly the same numerical conditions, but without the ambipolar term. This series is labeled as model D in \citet{Khomenko+etal2018} and it was used for the statistical comparison between the temperature and other parameters in the runs with and without ambipolar diffusion.

The three runs analyzed in this paper (\batt, \ambi, and \ambihall) are developed from the two last snapshots of the D and D-AD runs. Figure \ref{fig:NumExpData} shows schematically the development and the bifurcations of different runs in our numerical experiment. From the experiment D we obtain a new branch labeled \batt\  by running the dynamo simulation under exactly the same conditions for another 4.1 hours. The \ambi\ branch is developed from the latest D-AD snapshot and it is also run for the same amount of solar time. The \ambihall\ branch is obtained from the same D-AD snapshot after the Hall term is switched on in Eq. (\ref{eq:induction}). In this latter case, all three effects (battery, ambipolar and Hall) are present and acting simultaneously. The snapshots from these three branches are saved every 10 sec during the last 2 hours of each run, see Fig. \ref{fig:NumExpData}.

\begin{figure*}
\centering
\includegraphics[keepaspectratio,width=9cm]{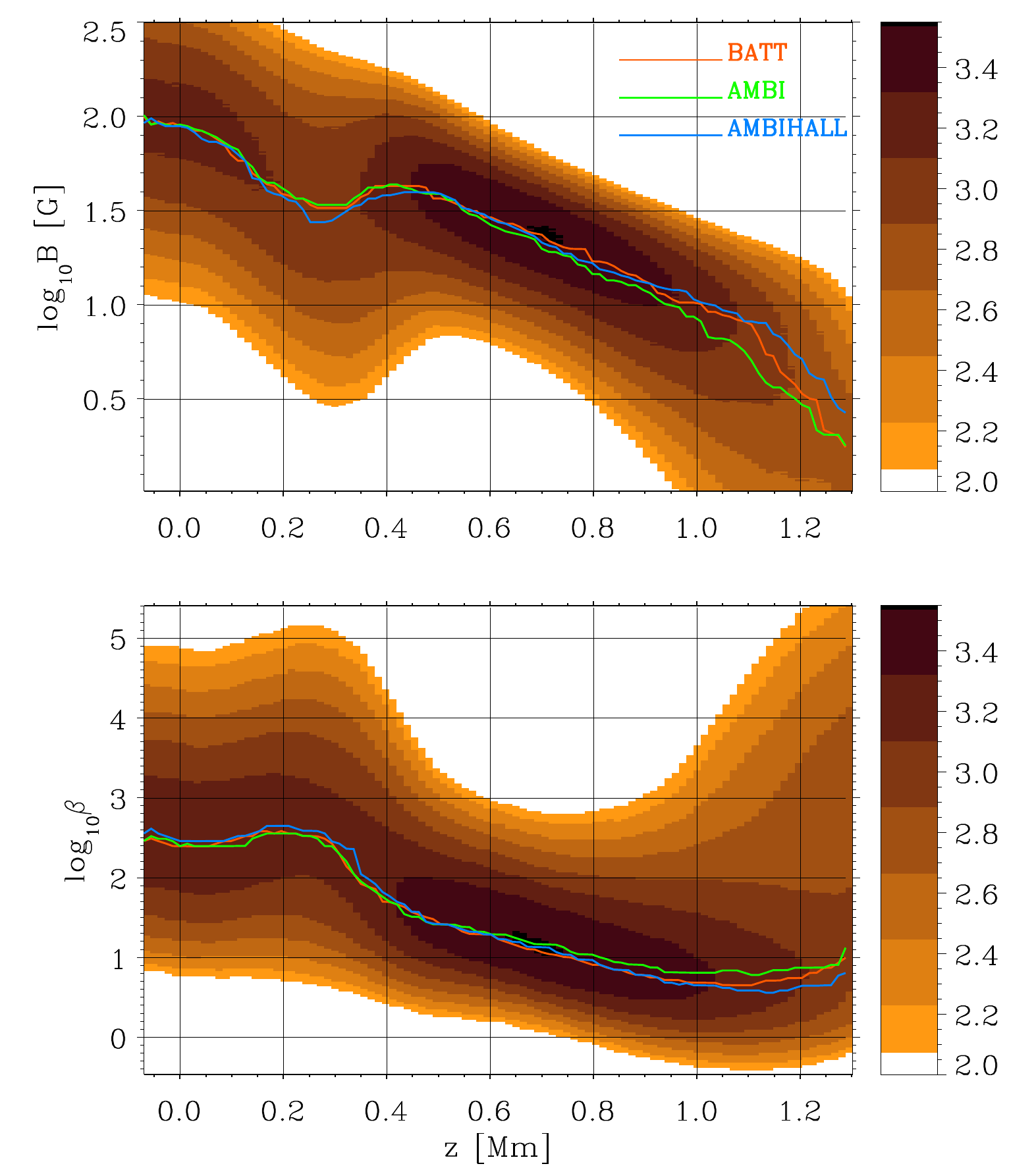}
\includegraphics[keepaspectratio,width=9cm]{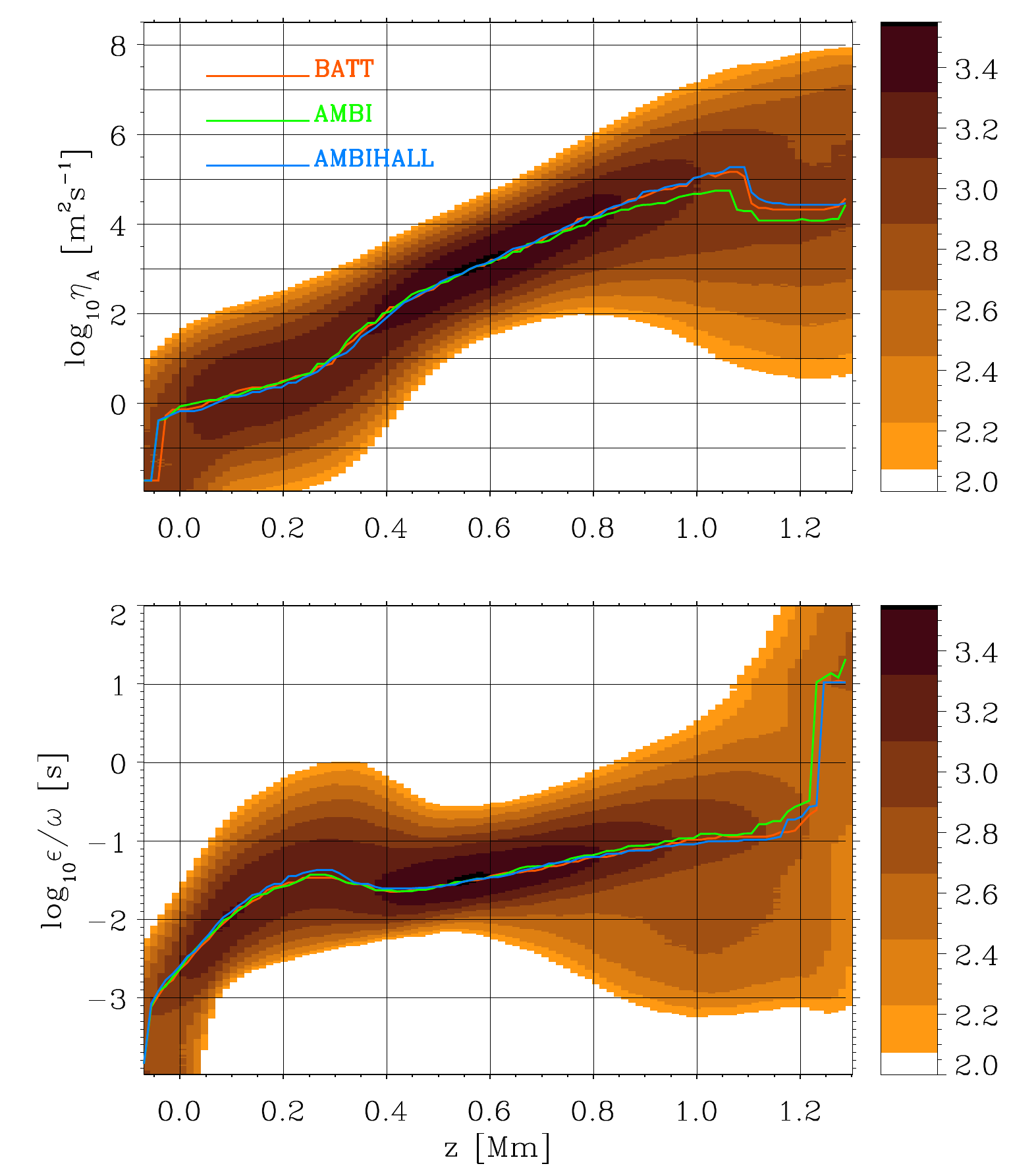}
\caption{\footnotesize Two-dimensional histograms showing the number of occurrences of a given value of a parameter, as a function of height in the \batt\ run. Darker colors mean larger probability of occurrence in logarithmic scale, indicated by the color bar. Red, green and blue lines follow the median value of the distribution at each height for the \batt, \ambi, and \ambihall\ runs, correspondingly. Top left: modulus of the magnetic field strength; top right: AD coefficient; bottom left: plasma-$\beta$; bottom right: Hall parameter $\epsilon/\omega$. }
\label{fig:distributions}
\end{figure*}

\subsection{Average properties of the simulations}

The simulations with/without certain effects are run for sufficiently long solar time to ensure the stationary state. Since these effects introduce perturbations, particular realizations of the convection pattern in all three simulations start to differ after about 10--15 min of solar time. Therefore, individual snapshots cannot be compared one to one, but can only be compared statistically. The parameters that influence most significantly the propagation and dissipation of energy throughout the simulation domain are the magnetic field strength, plasma-$\beta$, and the value of the ambipolar and the Hall coefficients.  Figure \ref{fig:distributions} gives two-dimensional histograms of these parameters. These histograms allow one to visualize the frequency of a given value of a parameter, as a function of height.

The two-dimensional distributions are similar in all three runs and cannot be appreciated in the format of the figure. The differences between the simulations can be seen by following the median of the distributions at each height. These median values are given by color lines in Figure \ref{fig:distributions}. The color lines overlap at heights below $\approx$0.6--0.8 Mm, and start to diverge at higher heights. 

The magnetic field strength curve in the \batt\ run is placed in between the one for the \ambi\ and \ambihall\ runs. The magnetic field is weaker in the \ambi\ model starting from $\approx$0.6 Mm upwards. This weakening was also observed in Paper I, see Figure 1 there. Such behavior is consistent with the fact that ambipolar diffusion acts by dissipating magnetic energy and converting it into thermal energy. The magnetic field strength can be expected to decrease, on average, in response to this action. The largest magnetic field is observed in the \ambihall\ model (blue line) at heights 0.9 Mm and above. It appears as, the additional currents created by the Hall effect in the weakly ionized chromospheric plasma redistribute the magnetic field and help intensifying magnetic structures in a way that stronger magnetic flux tubes reach higher in the atmosphere. \citet{Khodachenko+Zaitsev2002} suggested that the Hall effect can help to create intense flux tubes in the solar atmosphere, with the intensification happening from above (chromosphere) to below (photosphere). This effect needs further investigation. 

The distribution of plasma-$\beta$ in the simulations (Figure \ref{fig:distributions}, bottom left) reflects the one of the magnetic field strength. The median values of plasma-$\beta$ are the largest in the \ambi\ model and the lowest in the \ambihall\ model. There is only a small fraction of the volume where plasma-$\beta<1$, located above 0.8 Mm height. In most of the volume the gas pressure dominates over the magnetic pressure because the magnetic fields are relatively weak. The magnetic field in these models is structured in the form of low-lying closed loops, see Paper I. In many locations the field lines do not reach high, but return back to the surface, and this causes a drop of the average magnetic field strength and an increase of plasma-$\beta$ above  $\approx$ 0.8 Mm. 

The ambipolar diffusion coefficient is weakest for the \ambi\ simulation as a consequence of its weaker fields, and it is the largest for the \ambihall\ one (Figure \ref{fig:distributions}, upper right). Due to the closing of the loops and the drop of the field, the average value of the $\eta_A$ decreases in the chromosphere above 1.1 Mm. But nevertheless some significant part of the volume still shows values above 10$^6$ m$^2$s$^{-1}$. These locations with large ambipolar dissipation form elongated structures along the horizontal canopies connecting opposite polarity fields, see Paper I.

\begin{figure*}
\begin{center}
\includegraphics[width = 16cm]{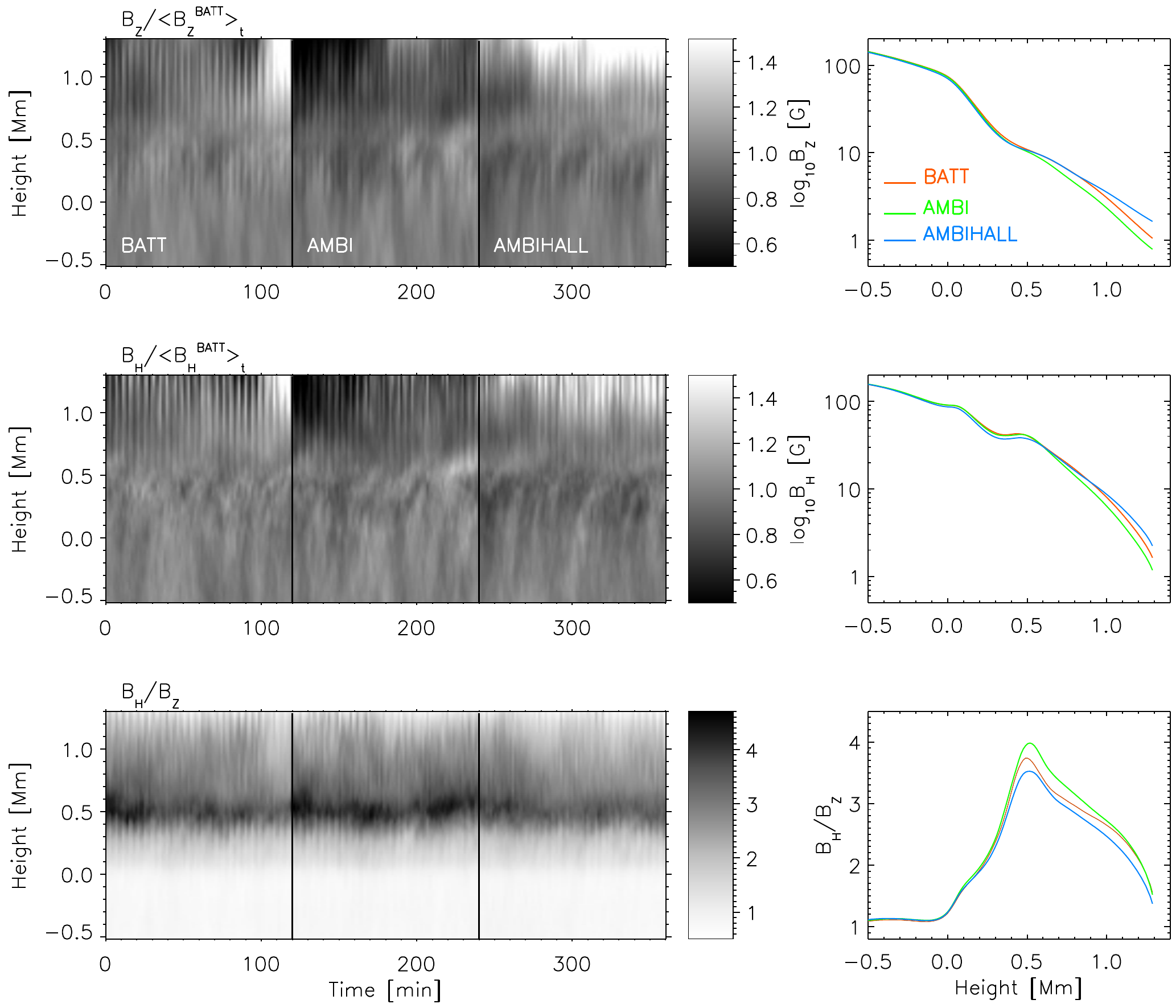}
\end{center}
\caption{\footnotesize  Top left: absolute value of the vertical magnetic field for the \batt, \ambi, and \ambihall\ models, averaged over the horizontal distance, as a function of time and height. The values for each model and each time are normalized to the time and horizontal average of $\langle |B_z| \rangle_{xyt}$ from the \batt\ model for a better visualization. Top right: time and horizontal average of the absolute value of the vertical magnetic field for the three models. Middle panels: same for the horizontal magnetic field $B_H$. Bottom panels: same for the ratio of the horizontal to vertical magnetic field components.} \label{fig:bhbz}
\end{figure*}

\begin{figure*}[!ht]
\centering
\includegraphics[keepaspectratio,width=16cm]{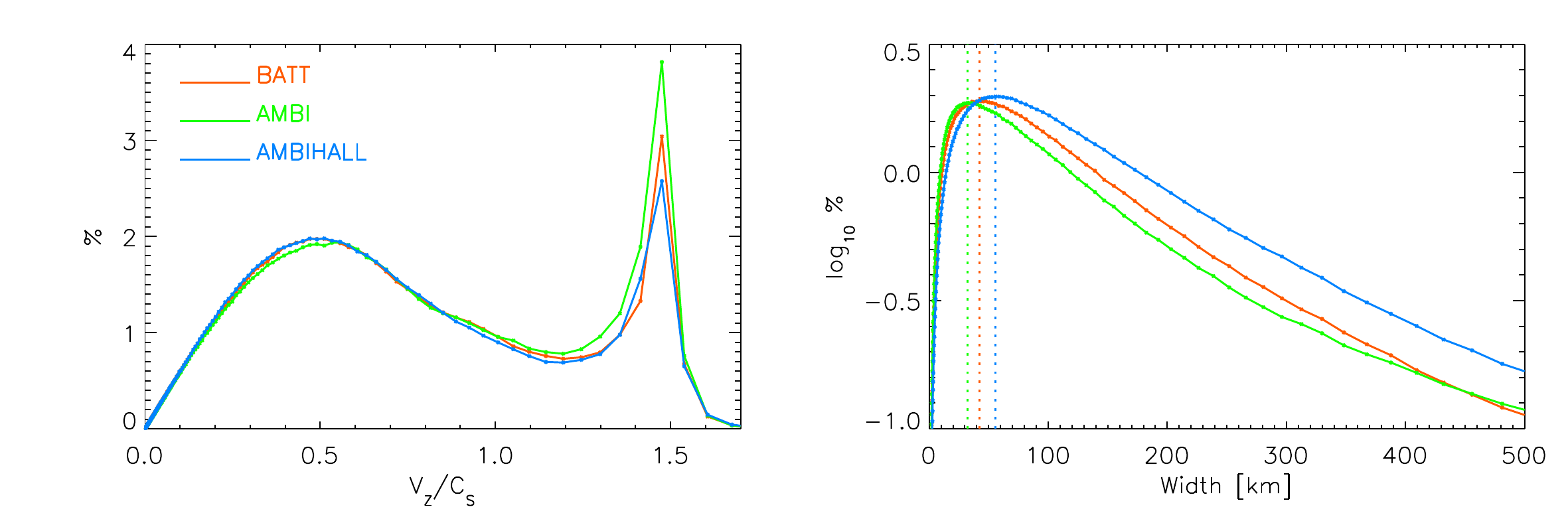}
\caption{\footnotesize Histograms of the vertical Mach number (left) and width of the current sheets (right) for the \batt, \ambi, and \ambihall\ models, at height 1.1 Mm above the photosphere. The size of bins at the horizontal axis in both cases is constant in log$_{10}$ scale. }\label{fig:mach_anch}
\end{figure*}


The Hall coefficient $\epsilon/\omega$ shown at the bottom right panel of Figure \ref{fig:distributions} is computed according to Eq. (\ref{eq:epsilon2}). The coefficient increases toward the upper part of the domain as a consequence of the weakening of the field and the relatively weak values of the Alfv\'en speed, $v_A$, in the chromosphere. The action of the Hall effect in our model benefits from such behavior of the magnetic field. For example, the height dependence of $\epsilon$ in our model can be compared with the one assumed in a more idealized study by \citet{Gonzalez-Morales+etal2019}, who used constant magnetic field strength with height and a VAL-C atmospheric model for thermodynamic quantities. In their model $\epsilon$ is only large in a window occupying a narrow height range in the photosphere, and it  exponentially drops below or above this layer. Here $\epsilon$ remains to be on the order of $10^{-1}$ over a significant part of the photosphere and rises in the chromosphere. An important fraction of the volume above 0.8 Mm has $\epsilon$ above 1, and even reaching 10 or higher. This means that the importance of the Hall effect spreads over temporal scales on the order of $1-10$ sec or even above.  Large values of the Hall parameter are expected to favor fast magneto-acoustic to Alfv\'en mode transformation \citep{Cally+Khomenko2015}, and therefore could help  bringing Alfv\'en waves to the chromosphere. These waves are expected to be dissipated throughout the ambipolar effect, closing the chain. This way, we have in our model the field generation by dynamo, its redistribution and possible creation of the Alfv\'en waves by the Hall effect, and  its dissipation by the ambipolar effect. The action of these three effects together has never been studied in a realistic context. Even our simulation lack dynamic description of hydrogen ionization, studying the interplay of these effects onto the chromospheric heating is an intriguing theoretical problem.


\subsubsection{ Amount of the emerged field}



The addition of the ambipolar diffusion to the simulations can modify the transport of magnetic flux, translating it from subphotospheric regions to the upper atmosphere, see \citet{Leake+Arber2006, Arber2007}. In a 2.5D idealized experiment, \citet{Leake+Arber2006} have shown that the emergence of magnetic flux can be greatly increased in partially ionized regions of the atmosphere and that the emerged magnetic flux is more diffused. When extending these experiments to 3D, \cite{Arber2007} concluded that less plasma is lifted to the corona compared to the 2.5D situation, and that no artificially cool layer is produced in 3D as the flux is emerged. Recently, \citet{Nobrega-Siverio+etal2020} performed realistic 2.5D simulations of flux emergence including radiation, nonequilibrium hydrogen ionization, and ambipolar diffusion. They have shown that the ambipolar diffusion does not significantly affect the amount of total unsigned emerged magnetic flux in their case, but it is important for the dissipation of shocks and the corresponding plasma heating.

While our experiments were not specifically designed to study the flux emergence, neither the simulation domain contains coronal regions, it is still interesting to check if the amount of the emerged flux is different when the ambipolar and Hall effects are added. There is no global or local flux emergence through the lower boundary in our simulations. The small-scale magnetic loops that are formed through the interaction of the magnetic field and convection inside the domain emerge through the convective layers to the surface. No one-to-one comparison of these magnetic loops and the associated currents can be performed in our three simulation series, as was done in a more controlled experiments mentioned above. Instead we followed the strategy by \citet{Leake+Arber2006} and computed the unsigned vertical flux, $\langle |B_z| \rangle_{xy}$, averaged in the horizontal direction. The results of this calculation are displayed in Figure \ref{fig:bhbz}, upper panel. The time-height plot at the upper left panel gives the evolution of $\langle |B_z| \rangle_{xy}$ for each of the three models in time. We have normalized this quantity to the time averaged value of $\langle |B_z| \rangle_{xyt}$ for the \batt\ run to better appreciate the differences between the simulations. It appears that, on average, no differences between the models are present in the deep layers below, approximately, 0.5 Mm. In the upper layers, there are both short-term and long-term variations of 
$\langle |B_z| \rangle_{xy}$, the former are related to wave periodicities of the chromospheric shocks. On the long-term, the \ambi\ simulation shows less  $\langle |B_z| \rangle_{xy}$ in the upper layers (darker regions), while the \ambihall\ shows the opposite (lighter regions). Therefore, it can be concluded that in experiment the ambipolar diffusion results in the less flux emergence to the chromospheric layers (or, alternatively, the magnetic flux has been dissipated after its emergence). 

The middle panel of Figure \ref{fig:bhbz} presents a similar analysis but for the horizontal component of the magnetic field, $B_H=\sqrt{B_x^2+B_y^2}$. It can be seen that the horizontal component of the magnetic field is also weaker in the \ambi\ case and is stronger in the \ambihall\ case, compared to the \batt\ case. Nevertheless, these nonideal effects do not affect the vertical and horizontal components in the same way. The bottom panel of  Figure \ref{fig:bhbz} shows that the ratio of the horizontal to vertical components is enhanced in the upper layers of the \ambi\ model, and is decreased for the \ambihall\ model. The magnetic fields that emerge to chromospheric layers are more vertical in the \ambihall\ case, and more horizontal in the \ambi\ case. Though these results cannot be directly compared to the studies of  \citet{Leake+Arber2006, Arber2007} and \citet{Nobrega-Siverio+etal2020}, they suggest that the role of the ambipolar and the Hall effect for the small-scale flux emergence in solar quiet regions needs a further deeper study.

\subsubsection{Properties of the current sheets}

nonideal terms can also affect the distribution of current sheets in the chromosphere. \cite{Arber2009} found that the presence of the ambipolar diffusion makes the magnetic field more force-free at chromospheric heights. Our simulations are far from a force-free situation in these chromospheric layers because of the weakness of the magnetic field. As can be seen in Figure \ref{fig:distributions} (bottom left), most of the domain in the chromosphere is in the large plasma $\beta$ regime, so that hydrodynamic forces are equally important as the magnetic ones. The ratio between the perpendicular to the total currents can be used as a measure of the  force-freeness of the field. This ratio is practically the same in our three models. Therefore, in the regime considered here, the ambipolar diffusion does not make structures in the chromosphere tangibly more force-free, on average. This can be due to the fact that the foot-points of the weak magnetic structures are constantly driven by the convective motions below the surface and the magnetic structures in the chromosphere do not have sufficient time to relax to an equilibrium situation. 

According to the work of \citet{Brandenburg+Zweibel1994}, ambipolar diffusion helps the formation of sharper current sheet structures and drives current sheet collapse. In order to check if there is any measurable difference in the width of the current sheets between the different models, we have evaluated the spatial sharpness of the currents for the three models in an approximate way. For that, we have computed $L_J=B/J\mu_0$ distributions as a direct measure of the gradients, for every height, spatial location, and time. Figure \ref{fig:mach_anch} (right panel) shows the distribution of $L_J$ computed this way at a height of 1.1 Mm above the photosphere.  It can be observed that the distributions for the three models are different. In all the cases, there is a long tail toward larger width values. However, the distribution is narrower and peaks at a lower value (32 km) for the \ambi\ model. The \batt\ model distribution peaks at 43 km and the \ambihall\ one peaks at 56 km. Only the simulation with the ambipolar diffusion shows indications for the narrower current sheets, in agreement with \citet{Brandenburg+Zweibel1994}. We have no explanation yet why adding the Hall effects acts in the opposite way. 

\subsubsection{Properties of shocks}

Ambipolar diffusion can influence the behavior of shocks and can act as an additional shock dissipation mechanism in the chromosphere \citep{Martnez-Sykora2020, Nobrega-Siverio+etal2020}. In our simulations we do not have a low plasma $\beta$ regime with slow-mode shocks propagating along well-defined and long-lived magnetic field lines. In the high-$\beta$ regime of our simulations, the most abundant shocks are fast shocks, that is, the usual hydrodynamic shocks. In order to check how important is the shocks contribution in our analysis, we have computed the vertical Mach number, $M=|v_z|/c_s$ as a function of height, horizontal distance, and time. Figure \ref{fig:mach_anch} (left panel) shows the distribution of $M$ at height 1.1 Mm for the three models. It appears that, (i) the high Mach numbers, $M>1$, are not very abundant in our simulations, and (ii) the high Mach number locations are slightly more abundant in the \ambi\ model and slightly less abundant in the \ambihall\ model, compared to the \batt\ model. This behavior may help understanding the power distribution of the fast wave components shown below in Figure \ref{fig:ffast}.

\begin{figure*}[!ht]
\centering
\includegraphics[keepaspectratio,width=16cm]{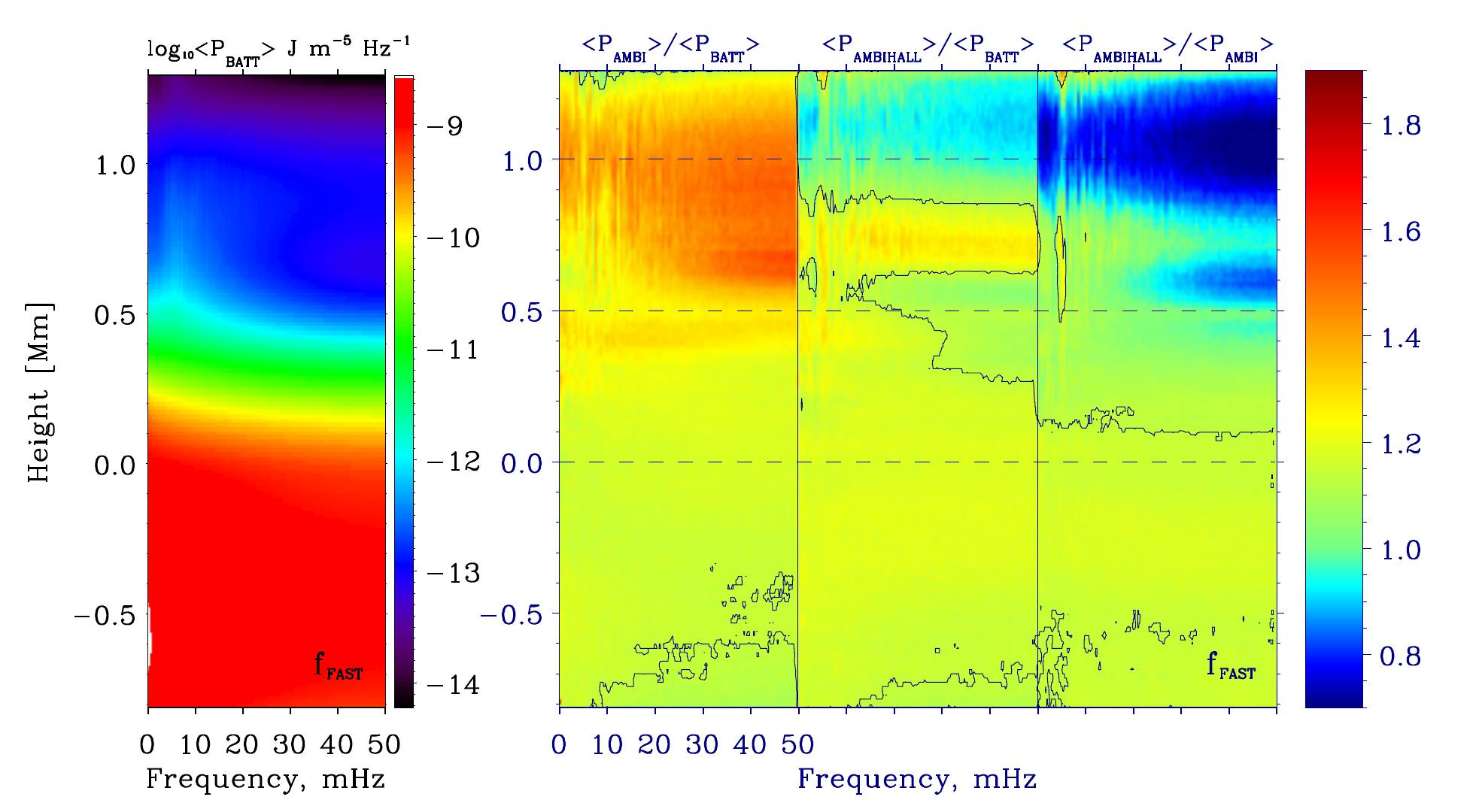}
\includegraphics[keepaspectratio,width=16cm]{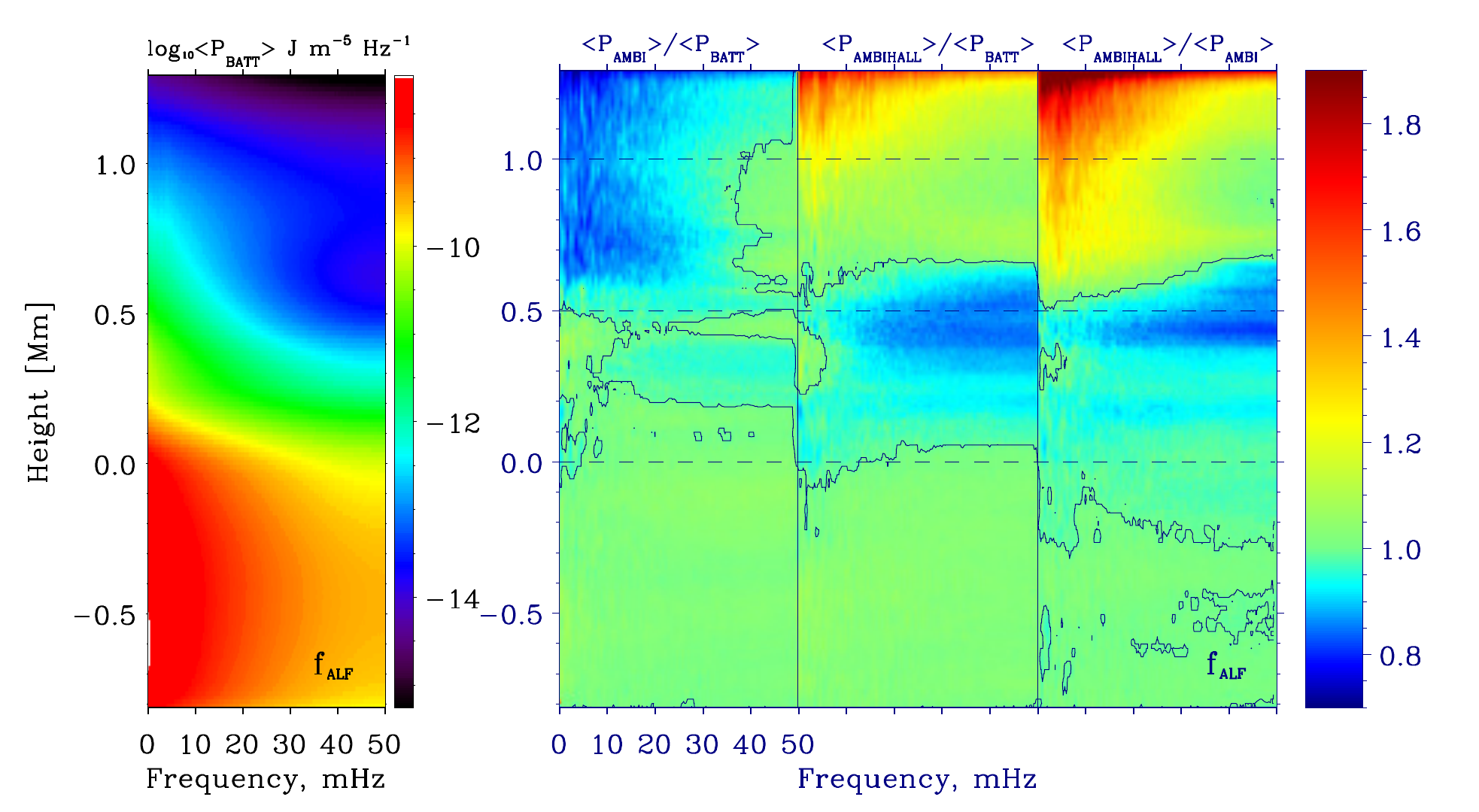}
\caption{\footnotesize Upper left: power spectra of $f_{\rm fast}$ as a function of height and frequency for the \batt\ run. For better visualization, the maps are scaled with a factor of $\rho_0$, the density of the unperturbed atmosphere. The power is shown in log$_{10}$ units of J m$ ^{-5}$ Hz$^{-1}$. Upper right panels: ratio of power between the \ambi\ and \batt, \ambihall\ and \batt, and \ambihall\ and \ambi\ runs.   Contours mark the locations where the power ratio is unity. Bottom panels: same, but for $f_{\rm alf}$.}\label{fig:ffast}
\end{figure*}

\begin{figure*}[!ht]
\centering
\includegraphics[keepaspectratio,width=16cm]{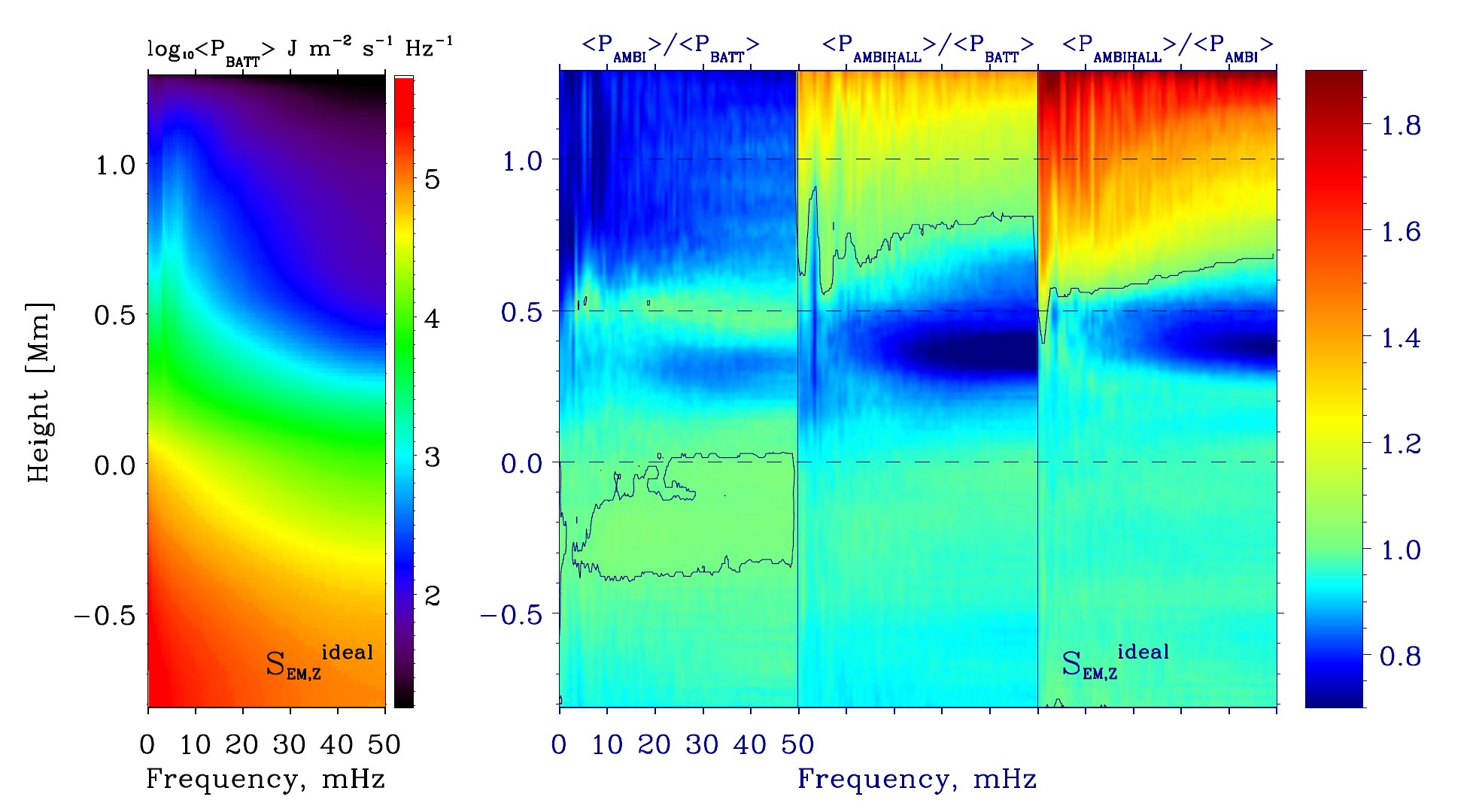}
\caption{\footnotesize Left: power spectra of $\mathbf{S}_{\rm EM, z}^{\rm ideal}$ (ideal part) as a function of height and frequency for the \batt\ run. The power is shown in log$_{10}$ units of J m$ ^{-2}$s$^{-1}$ Hz$^{-1}$. Upper right panels: ratio of power between the \ambi\ and \batt, \ambihall\ and \batt, and \ambihall\ and \ambi\ runs.   Contours mark the locations where the power ratio is unity. }\label{fig:fmagz}
\end{figure*}

\section{Statistical Fourier analysis}

The 10 sec cadence and almost 2 hours length of our simulated series allow us to study perturbations in a wide range of frequencies. Following Paper I strategy, we study separately the incompressible and compressible perturbations. For that we compute the following quantities, as suggested by \citet{Cally2017}:
\begin{equation} \label{eq:falf}
\mathrm{f}_{\rm alf} = \hat{e}_\parallel \,\mathbf{\cdot \nabla \times} \mathbf{v},
\end{equation}
\begin{equation}\label{eq:ffast}
\mathrm{f}_{\rm fast} =\mathbf{\nabla\, \cdot} \left(\mathbf{v} -  \hat{e}_\parallel \,v_{\parallel}\right) = \mathbf{\nabla\cdot v}_\perp.
\end{equation}
%
%
Here, $v_{\parallel} =  \hat{e}_\parallel \cdot \mathbf{v}$ and $\mathbf{v}_\perp=\mathbf{v} -  \hat{e}_\parallel \,v_{\parallel}$ are the velocity components parallel and perpendicular to the magnetic field, and $ \hat{e}_\parallel$ is the field-aligned unit vector. The parameter $\mathrm{f}_{\rm alf}$ describes incompressible perturbations propagating along the field lines, that can be associated with Alf\'en waves, independently of the value of plasma $\beta$. The $\mathrm{f}_{\rm fast}$  describes the compressible perturbations in velocity perpendicular to the magnetic field lines. In the low-$\beta$ plasma, $\mathrm{f}_{\rm fast}$  would select fast magneto-acoustic waves. In our case, the plasma $\beta$ is not low enough to maintain anisotropy in the compressible perturbations parallel and perpendicular to the magnetic field lines. In most of the volume $\mathrm{f}_{\rm fast}$ would just select a component of the compressible (magneto)-acoustic perturbation. 

Compared to Paper I, we do not introduce $\mathrm{f}_{\rm long}$. The behavior of this quantity (which is a compressible perturbation in velocity parallel to the magnetic field) is only expected to be different from that of $\mathrm{f}_{\rm fast}$ if the plasma beta is sufficiently low. The behavior of $\mathrm{f}_{\rm long}$ and $\mathrm{f}_{\rm fast}$ was indeed similar in the dynamo simulations analyzed in Paper I.

In principle, the decomposition above into compressible and incompressible perturbations can be affected by shocks, due to the sharp nature of the shock fronts. However, as discussed above, under the circumstances of our model (high-$\beta$) most of the shocks are fast shocks. The slow waves are near-transverse and not very compressive. According to recent results by \citet{Pennicott+Cally2019}, who studied the behavior of shocks as they cross the $\beta=1$ layer, slow shocks are very weak both in compression ratio and in Mach number. The slow shocks can produce considerable magnetic field shear and change the field direction. Mode conversion still happens at $\beta=1$ as the shock propagates, splitting the shock into slow and fast parts. The fast part remains a shock, but the slow part is smoothed out. Because of that, and also because the high Mach number locations are not very abundant (see Figure \ref{fig:mach_anch}), we believe that the decomposition we perform here between the Alfv\'en and fast components is not significantly affected by the presence of the sharp discontinuities due to shocks. 

Following the method established in Paper I, we have computed the power spectra of the $\mathrm{f}_{\rm alf}$  and $\mathrm{f}_{\rm fast}$ perturbations and have produced maps where these spectra are averaged over the horizontal coordinates. These power maps have been obtained as a function of height and frequency. The ratio between the power maps between pairs of simulations gives the maps of the power deficit or excess,
\begin{equation}\label{eq:Fratio}
F_i(z,\nu) = \frac{\left<\left|\mathrm{FFT}\left(f_i \right)_\mathrm{R1}\right|^2\right>_{x,y}}{\left<\left|\mathrm{FFT}\left(f_i \right)_\mathrm{R2}\right|^2\right>_{x,y}},
\end{equation}
where  the quantity $f_i = \{ \mathrm{f}_{\rm alf}, \mathrm{f}_{\rm fast} \}$ indicates the projection and the superscript (R1,R2) = \{\batt, \ambi, \ambihall\} indicates the simulation. The power maps computed this way are shown in Figure \ref{fig:ffast} for  $\mathrm{f}_{\rm fast}$  (top) and  $\mathrm{f}_{\rm fast}$ (bottom).

The left panels of Figure \ref{fig:ffast} provide the power spectrum computed for the \batt\ run. The spectrum of $\mathrm{f}_{\rm fast}$ shows a typical distribution, proper for solar convection and oscillations. The spectrum shows a peak for frequencies around  3--5 mHz, typical for solar oscillations, and a smooth decay in power toward the higher frequencies. The power ratio maps given on the right, demonstrate differences between the three simulations in layers above 300--400 km. The ratio $\langle P \rangle_{\rm AMBI}/\langle P \rangle_{\rm BATT}$, shows an  excess of power of $\mathrm{f}_{\rm fast}$  in the \ambi\ run of up to 40\%. This excess is especially prominent at higher frequencies, where it extends to lower layers. This power ratio map can be compared to the one in Paper I, see Figure 5 there, where a shorter series was used with two times lower cadence of snapshots saving. The data from Paper I showed a power excess around 1 Mm and some power depression around 0.6 Mm. The new simulation series shows a somewhat larger power excess at the same frequency range as in  Paper I (below 25 mHz), and almost negligible power depression at 0.6 Mm. By constructing power maps for the first hour and the second hour of our simulations separately, with a cadence of 20 sec, we have confirmed that these differences are due to statistical fluctuations in the datasets. However, the general trend is preserved and is much better  resolved in frequency in the new longer series of simulations used in this work: the presence of ambipolar diffusion results in an excess of power of $\mathrm{f}_{\rm fast}$ compressible fluctuations in the chromosphere, especially prominent at higher frequencies above 25 mHz.

This excess in the power can be linked to the Mach number distributions in Figure \ref{fig:mach_anch} above, which shows that the \ambi\ simulation produces slightly more abundant shocks. The fast waves represented by $f_{\rm fast}$ are those that shock at chromospheric heights, and according to  Figure \ref{fig:ffast} they have larger amplitudes at high frequencies in the \ambi\ case. These high-frequency waves steepen to shocks quicker, resulting in more abundant locations with large $M$ in Figure \ref{fig:mach_anch}. 

The power ratio $\langle P \rangle_{\rm AMBIHALL}/\langle P \rangle_{\rm BATT}$ shows almost no prominent features, just a weak power excess around 0.6 Mm and a weak depression at higher layers. The \ambihall\ simulation contains the action of both effects together and it is difficult to separate their action. The action of the Hall effect alone can be better checked in the power ratio map $\langle P \rangle_{\rm AMBIHALL}/\langle P \rangle_{\rm AMBI}$ (rightmost image). One can observe that the Hall effect removes the power of $\mathrm{f}_{\rm fast}$ fluctuations around 1 Mm. The power depression increases with frequency, and reaches about 30--40 \% at the maximum frequency resolved in our simulations, 50 mHz.  As mentioned above, such behavior can be either because the Hall effect modifies the structure of the magnetic fields in the model, or (and)  because it affects the efficiency of the mode transformation redistributing the power of different oscillation modes as a function of frequency and magnetic field topology. 

The power map of $\mathrm{f}_{\rm alf}$ (bottom left panel of  Figure \ref{fig:ffast}) shows some indication of a peak corresponding to 3 mHz oscillations, and a much steeper decay of power with frequency compared to  $\mathrm{f}_{\rm fast}$. The power ratio $\langle P \rangle_{\rm AMBI}/\langle P \rangle_{\rm BATT}$, shows about 20--30\% decrease of power of the Alfv\'enic incompressible fluctuations at layers above 600 km and frequencies below 30 mHz. A similar behavior was observed in Paper I, with slightly larger values of the depression. It is surprising that the power depression does not extend to higher frequencies, as one might expect from general considerations. This needs to be investigated further. 

The power ratio $\langle P \rangle_{\rm AMBIHALL}/\langle P \rangle_{\rm BATT}$ computed for $\mathrm{f}_{\rm alf}$ demonstrates that the action of both ambipolar and the Hall effects together results in 40--60\% power excess above 1 Mm height. The origin of this excess is the action of the Hall effect, as can be seen at the $\langle P \rangle_{\rm AMBIHALL}/\langle P \rangle_{\rm AMBI}$ map at the rightmost image. The Hall effect produces almost twice larger power of $\mathrm{f}_{\rm alf}$ in the upper layers of the domain. This increase in the power is frequency-dependent, and its maximum falls at lower frequencies, where it extends to lower layers as well. This, again, is a counter-intuitive result. One may expect the nonideal effects become more important for smaller scales and higher frequencies. The theoretical study of the mode transformation induced by the Hall effect in \citet{Gonzalez-Morales+etal2019} revealed that the higher the frequency, the larger the power of Alfv\'en waves produced through this effect. The situation in realistic simulations is significantly more complex than in the idealized study by \citet{Gonzalez-Morales+etal2019} since the magnetic structures also change dynamically on different scales moved by convection. We address these questions in the Section \ref{sect:hifreq} below.

\subsection{Poynting flux}

Following Paper I, we consider the Poynting flux reaching the upper layers of the simulation domain. The Poynting flux,
\begin{equation}\label{eq:poynting_all}
\mathbf{S}= -\frac{(\mathbf{v} \times \mathbf{B}) \times \mathbf{B}}{\mu_0} + \eta_H  \frac{|B|^2\mathbf{J}_\perp-\nabla p_\mathrm{e} \times \mathbf{B}}{|B|\mu_0} -\frac{\mathbf{B} \times\left(\eta_A \mathbf{J}_\perp \right)}{\mu_0},
\end{equation}
can be split into the ideal and nonideal parts. The ideal part is the first term in the equation above,
\begin{equation} \label{eq:poynting-ideal}
\mathbf{S}_{\rm EM}^{\rm ideal}=-\frac{(\mathbf{v}\times\mathbf{B})\times\mathbf{B}}{\mu_0}. 
\end{equation}
The remaining terms are the contributors from the Hall, battery and ambipolar effects. The Poynting flux $\mathbf{S}$ in the three simulations can be assumed, on average, the same, since the simulations are alike realizations of convection with some statistical variations. The nonideal effects are negligible at the bottom of the domain in the subsurface layers, and $\mathbf{S} \approx \mathbf{S}_{\rm EM}^{\rm ideal}$, and is approximately the same in all there cases. 

In the surface layers, the nonideal effects play a role. Since, on average, $\mathbf{S}$ is expected to be very similar in the three simulations because of the arguments above, the ideal part, $\mathbf{S}_{\rm EM}^{\rm ideal}$, can vary, depending on the sign and magnitude of the contribution from the nonideal terms.  Therefore, differences in $\mathbf{S}_{\rm EM}^{\rm ideal}$ can appear between the \batt, \ambi, and \ambihall\ simulations. The contribution of the battery term, present in all three runs, is very small. The remaining contributions, ambipolar and Hall ones, can either remove or add to the total Poynting flux. The ambipolar effect is dissipative, as follows from the energy conservation equation. The presence of the ambipolar term should therefore decrease the ideal Poynting flux, $\mathbf{S}_{\rm EM}^{\rm ideal}$. The Hall contribution is not dissipative as it does not make it into the energy equation. The Hall effect distributes the magnetic Poynting flux, but does not dissipate it into heat. However, the redistribution can affect $\mathbf{S}_{\rm EM}^{\rm ideal}$ reaching the upper layers. 

Figure \ref{fig:fmagz} shows the power map and the power ratio maps of the vertical component of $\mathbf{S}_{\rm EM}^{\rm ideal}$ in the same format as in Figure \ref{fig:ffast} above. The power map (left panel) shows the presence of an oscillation peak at 3--5 mHz, and a drop of power toward the higher frequencies. The power map ratio $\langle P \rangle_{\rm AMBI}/\langle P \rangle_{\rm BATT}$ reveals that there is a strong absorption of the Poynting flux in the simulation with the ambipolar effect relative to the \batt\ one, similar to what was already obtained in Paper I. The amount of the Poynting flux absorption is somewhat less than in Paper I, reaching 30--40\%. The $\langle P \rangle_{\rm AMBIHALL}/\langle P \rangle_{\rm BATT}$ power ratio shows a layer with absorption of the flux around 400--500 km, with the amount  of absorption increasing with frequency. In the upper layers, this figure reveals some 20\% excess of the flux, rather constant with frequency. The $\langle P \rangle_{\rm AMBIHALL}/\langle P \rangle_{\rm AMBI}$ map shows that the reason of the emission at the upper layers, and absorption at the lower layers is the action of the Hall effect. Relative to \ambi\ simulation, the \ambihall\ one produces about twice larger Poynting flux to the upper layers. 

\begin{figure}
\begin{center}
\includegraphics[width = 8cm]{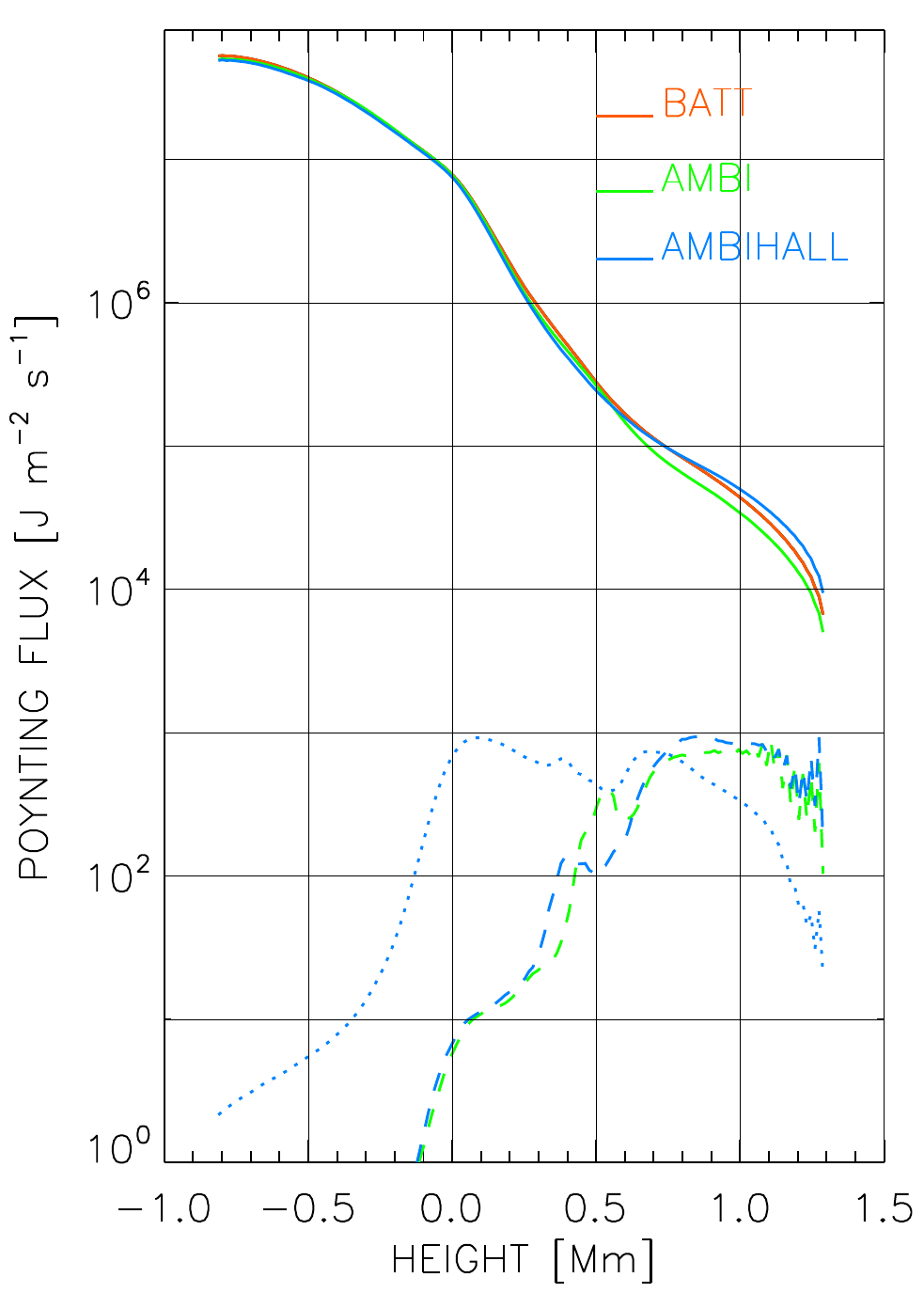}
\end{center}
\caption{\footnotesize Solid lines: total magnitude of the vertical component of the ideal Poynting flux, $\mathbf{S}_{\rm EM}^{\rm ideal}$, as a function of height for the \batt\ (red), \ambi\ (green), and \ambihall\ (blue) cases. Dashed lines show the total magnitude of the vertical component of the ambipolar contribution from Eq.\ref{eq:poynting_all} for the \ambi\ model (green) and \ambihall\ model (blue). Dotted blue line shows the total magnitude of the vertical component of the Hall contribution.} \label{fig:poynting_total}
\end{figure}

Figure \ref{fig:poynting_total} shows the total magnitude of the vertical component of the ideal Poynting flux, $\mathbf{S}_{\rm EM}^{\rm ideal}$, as a function of height, computed by adding the contribution by all frequencies for the three models. In addition, it gives the total magnitude of the vertical component of the ambipolar and the Hall contributions (see Eq. \ref{eq:poynting_all}) in the models where these terms were switched on. The ambipolar contribution (dashed lines) is maximum between 0.5 and 1.3 Mm, and it drops to very low values in the deep layers because of the relative absence of neutrals in a highly-collisional medium. It also slightly decreases in the upper layers because of the low values of the magnetic field, and the increase of temperature causing a decrease in the neutral fraction. The average magnitude of the ambipolar contribution is negative (the figure shows Fourier amplitude). The Hall contribution (dotted line) is large between 0 and 1.2 Mm, and this term takes both positive and negative values, with no clear preference. There is no significant flux due to ambipolar and Hall contributions through the upper boundary and these contributions are completely negligible at the low boundary and do not affect the flux transported through the lower boundary of our domain.



\begin{figure}
\centering
\includegraphics[keepaspectratio,width=9cm]{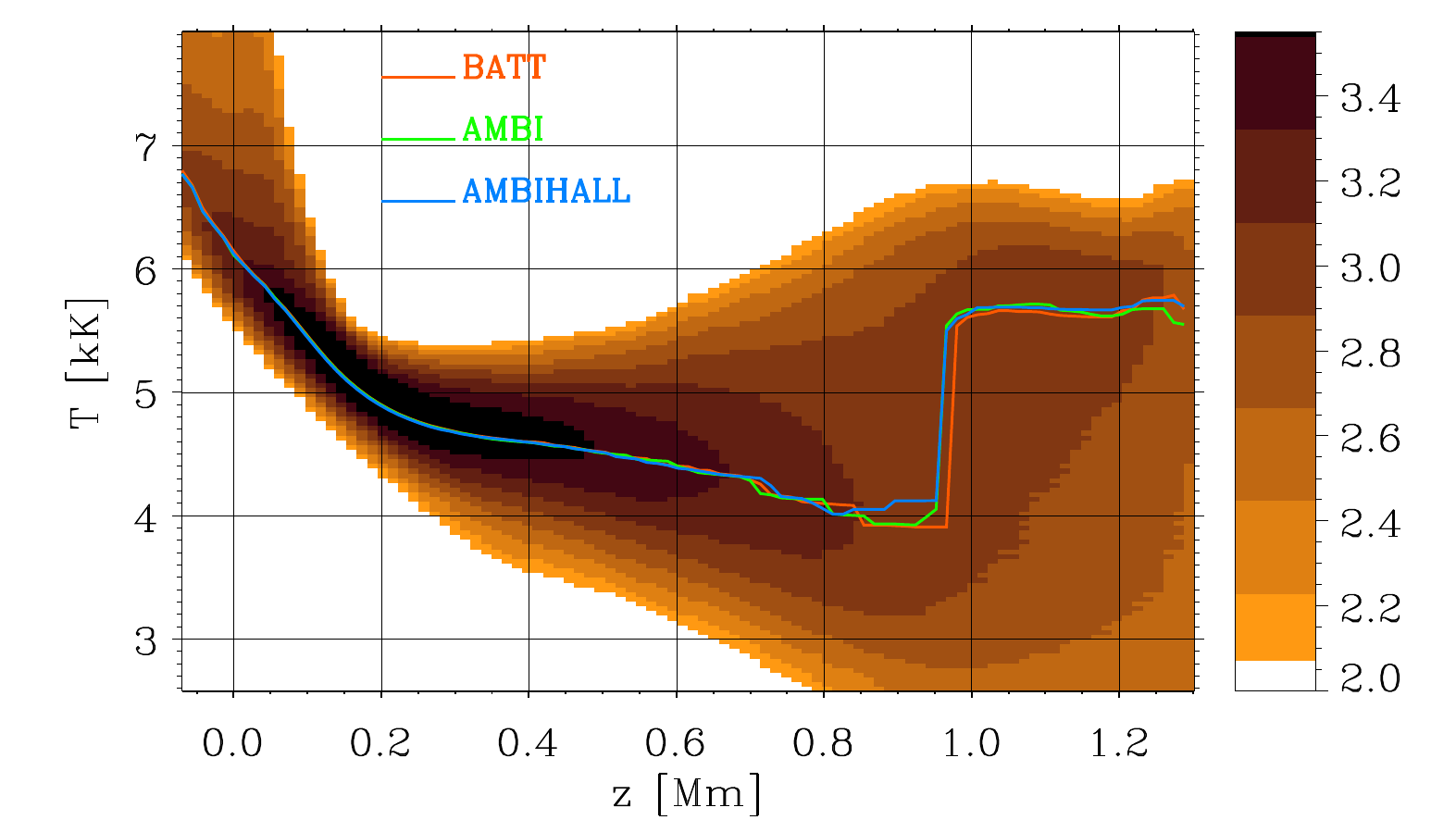}
\caption{\footnotesize  Two-dimensional histogram showing the number of occurrences of a given value of temperature, as a function of height in the \batt\ run. Darker colors mean larger probability of occurrence in logarithmic scale, indicated by the color bar. Red, green and blue lines follow the median value of the distribution at each height for the \batt, \ambi, and \ambihall\ runs, correspondingly. }\label{fig:temperature}
\end{figure}

\subsection{Effects on temperature}

The magnetic energy deposited in the atmosphere has some effect on the average temperature distribution. Figure \ref{fig:temperature} shows the two-dimensional histogram of the temperature in the \batt\ simulations. The distributions are rather similar for the other two runs, as presented in this format. In order to see the differences, we have computed the median of the distributions at each height, the same to what was done in Figure \ref{fig:distributions}. These median values are plotted as color lines over the two-dimensional histogram. It can be observed that the \ambi\ simulation (greed line) has larger temperatures than the \batt\ one at heights of 0.9--1 Mm, similarly to the result obtained in Paper I. The \ambihall\ simulation shows larger temperatures already at the deeper layers, between 0.7 and 1 Mm.

\begin{figure*}[!ht]
\centering
\includegraphics[keepaspectratio,width=16cm]{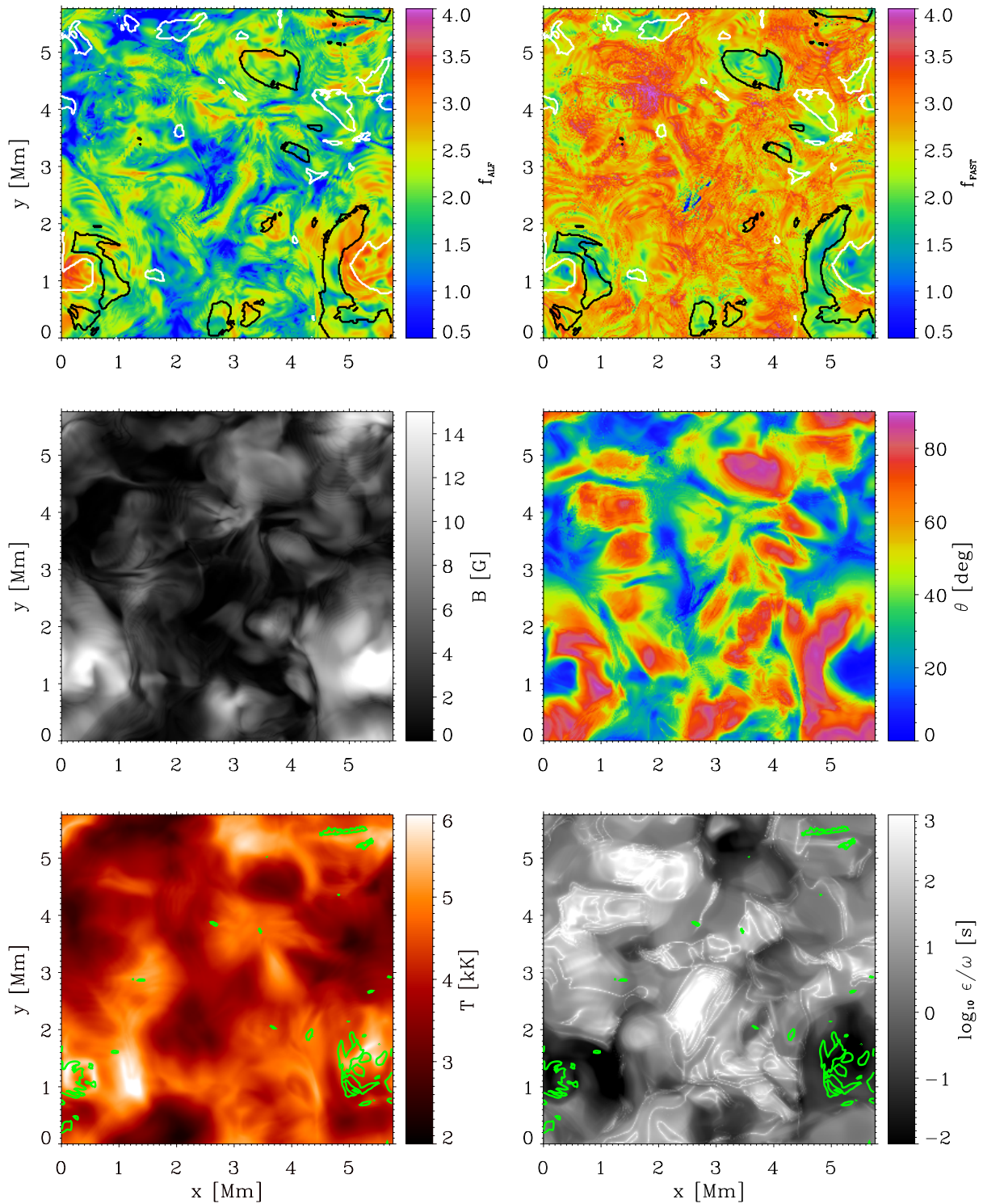}
\caption{\footnotesize  Images of log$_{10}$ of integrated high-frequency power of $\mathrm{f}_{\rm alf}$ (upper left); the power of $\mathrm{f}_{\rm fast}$ (upper right); magnetic field strength (middle left); its inclination (middle right); temperature (bottom left) and the Hall parameter, log$_{10}\epsilon/\omega$ (bottom right) at height 1.2 Mm in the \ambihall\ simulation. Contours at the upper panels indicate locations with plasma-$\beta$ is below 10 and the magnetic field inclination is $\theta<20$ degrees (white lines) or $\theta>70$ degrees (black lines). Contours at the bottom panels indicate extreme values of log$_{10}\mathcal{P}_{\mathrm{f}_{\rm alf}}$=2.9.}\label{fig:maps-hifreq-ambihall}
\end{figure*}

\section{High-frequency waves}
\label{sect:hifreq}

In order to understand the results above, it is necessary to study correlations between the properties of waves (or perturbations) and the structure of the magnetic elements and the granulation. This cannot be done, however, for waves at the frequency range where the power is maximum, 3--5 mHz. The typical life time of granulation pattern is 8--10 min, comparable to the period of such waves (3--5 min). The structures do not live enough time to meaningfully study the propagation of long-period waves through them.

Since the effects of  power absorption and emission are present at all frequencies, we can study the correlations for the high-frequency waves, ``freezing'' the granulation pattern. For that, we have computed the total high-frequency power, $\mathcal{P}_i$, Eq. (\ref{eq:P_i}). We have taken the last 50 min of each of the three simulation series, and split this interval of time into $n=25$ subintervals of 2 min duration. We have computed the power spectra of the $f$-quantities, and of the Poynting flux in each subinterval, and averaged those spectra in the complete frequency range from 8.3 mHz to 50 mHz, 

\begin{equation}\label{eq:P_i}
\mathcal{P}_i(x,y,z, t_n)=\sum_{\nu} \bigg| \mathrm{FFT}\Big[f_i(x,y,z,t_n:t_n+120\, {\rm sec})\Big]\bigg|^2,
\end{equation}
where $f_i = \{ \mathrm{f}_{\rm alf}, \mathrm{f}_{\rm fast},\mathbf{S}_{\rm EM}^{\rm ideal} \}$, and  $t_n=4200 +n\times 120$ sec. Each interval contains 12 snapshots, which is a compromise of having sufficiently long series, needed for statistically valid analysis, and having it sufficiently short in order to freeze the granulation. 

Simultaneously, we have computed averaged quantities of interest, such as average magnetic field strength, its inclination, temperature, values of the ambipolar heating, and Hall coefficient as 
\begin{equation}
\mathcal{S}_i(x,y,z, t_n)=\langle g_i(x,y,z,t_n:t_n+120\, {\rm sec}) \rangle,
\end{equation}

with $g_i=\{ B,\theta,T,Q_{\rm AMB},\epsilon/w \}$. We used averages over the same intervals for which the power spectra were computed. In the following, we study the correlations between $\mathcal{P}_i(x,y,z, t_n)$ and $\mathcal{S}_i(x,y,z, t_n)$.

Figure \ref{fig:maps-hifreq-ambihall} illustrates the distribution of several quantities over the surface at height 1.2 Mm in the chromosphere. It can be appreciated that the distribution of power of $\mathrm{f}_{\rm alf}$  and $\mathrm{f}_{\rm fast}$ is complementary to each other. Where the power of one quantity is larger, the other is smaller. Both distributions correlate with the magnetic field topology, see the contours of plasma-$\beta$ over the upper two panels.  Stronger magnetic field structures, with plasma-$\beta$ below 10, mostly have larger power of $\mathrm{f}_{\rm alf}$ generated at their centers. This is the case, for example, of the structures located around $\{x,y\}=\{0.2,1\}$ Mm, $\{5.5, 1.2\}$ Mm, and $\{3.5, 4.7\}$ Mm. The power of $\mathrm{f}_{\rm alf}$ is maximum for vertical magnetic fields (white contours). The situation is the opposite for $\mathrm{f}_{\rm fast}$, the power of compressible perturbations is maximum in regions with weak magnetic field and it does not apparently correlate with the magnetic field inclination. 

The temperature variations (Fig. \ref{fig:maps-hifreq-ambihall}, bottom left) demonstrate that the temperature is maximum at locations with maximum field strength and maximum power of $\mathrm{f}_{\rm alf}$. These locations are highlighted with the green contours. The correlations between the power in $f$-quantities and the Hall parameter are more difficult to establish. One may expect from theoretical grounds \citep{Cally+Khomenko2015, Gonzalez-Morales+etal2019} that the power of Alfv\'enic fluctuations, $\mathrm{f}_{\rm alf}$, generated after the Hall-induced transformation is maximum for the vertical fields. Fig. \ref{fig:maps-hifreq-ambihall}, bottom right shows the opposite effect: the Hall parameter is minimum where the power of $\mathrm{f}_{\rm alf}$ is maximum (see green contours). The Hall parameter has a strong dependence on temperature through the electron number density. At locations with large temperatures, the electron number density increases because the plasma becomes more ionized, and $\epsilon$ becomes lower, according to Eq. (\ref{eq:epsilon1}). Therefore, it is hard to conclude if the Hall-induced transformation plays a role in our simulations, but we cannot rule it out either just from the information present in Fig. \ref{fig:maps-hifreq-ambihall}.
As has been already seen, the \ambihall\ simulation contains stronger fields and larger fraction of areas with low plasma-$\beta$. Since, according to Fig. \ref{fig:maps-hifreq-ambihall}, these stronger fields are associated with a larger power of $\mathrm{f}_{\rm alf}$, we conclude that this might be responsible for the larger Poynting flux to the upper layers in the \ambihall\ simulations, as present in Fig. \ref{fig:fmagz}.

\begin{figure*}
\centering
\includegraphics[keepaspectratio,width=16cm]{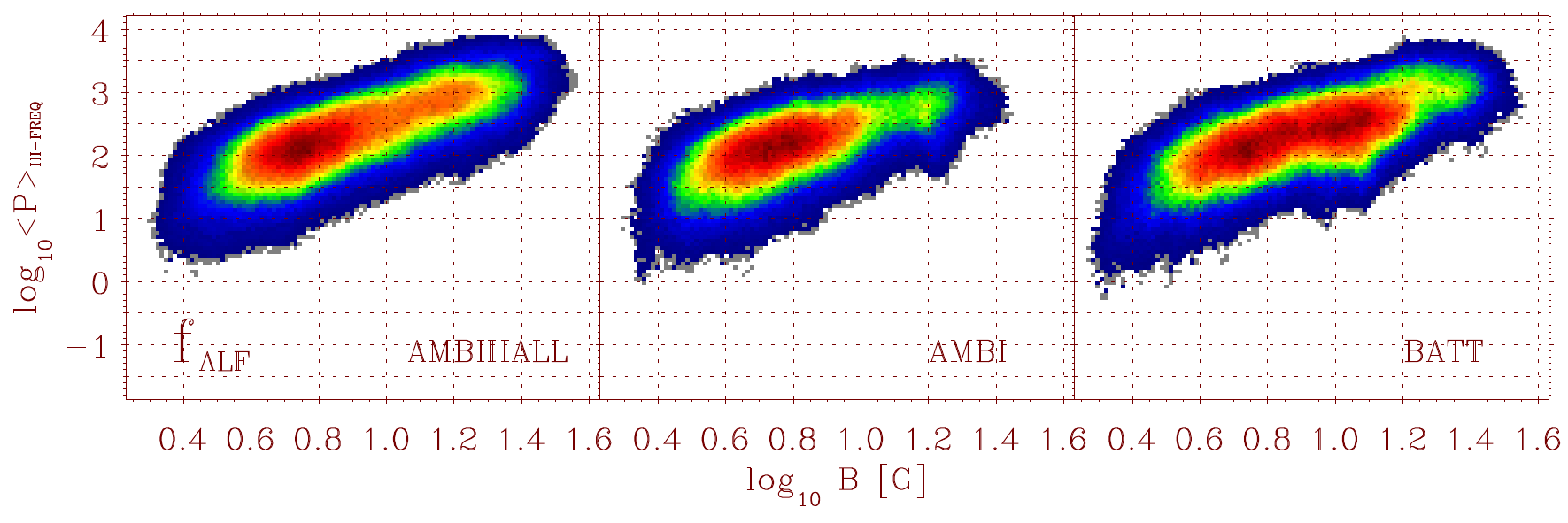}
\includegraphics[keepaspectratio,width=16cm]{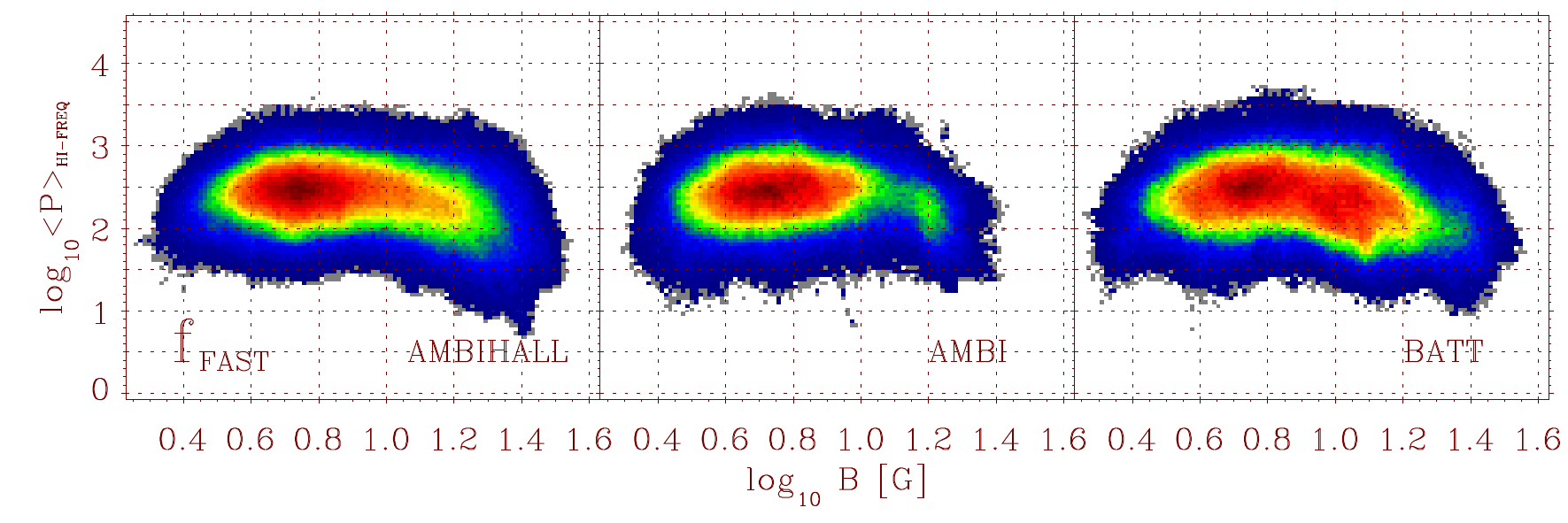}
\includegraphics[keepaspectratio,width=16cm]{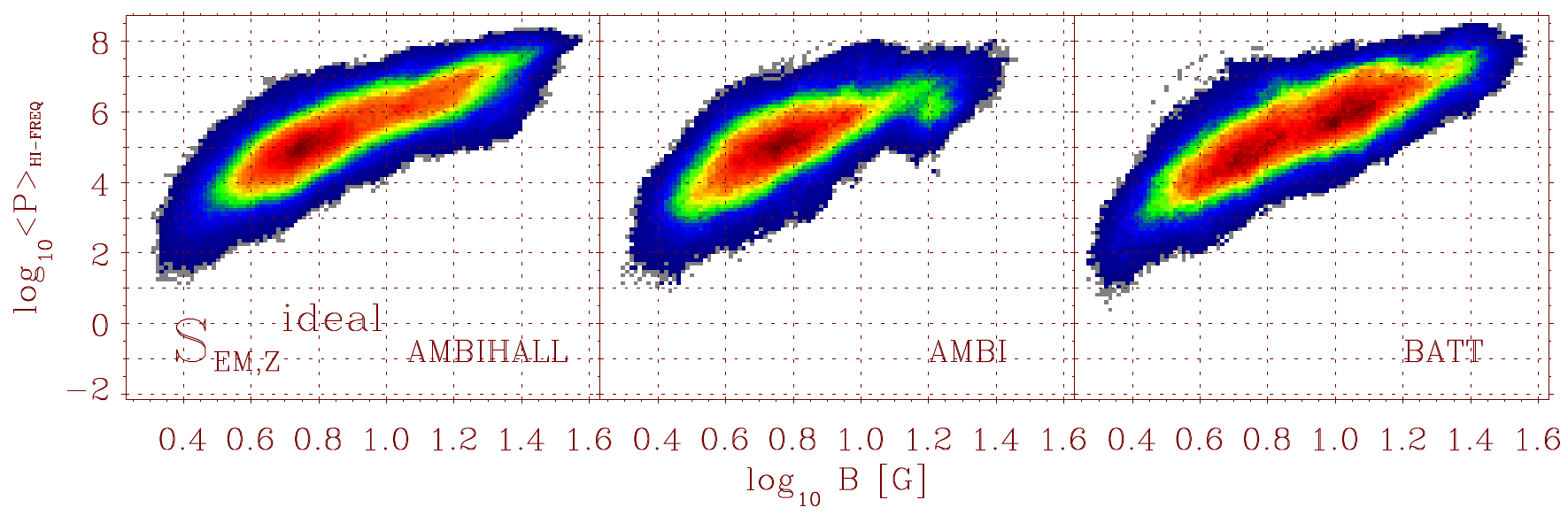}
\caption{\footnotesize  Two-dimensional histograms showing the dependence on magnetic field strength, $B$, of the power of $\mathrm{f}_{\rm alf}$ (top), $\mathrm{f}_{\rm fast}$ (middle), and $\mathbf{S}_{\rm EM,z}^{\rm ideal}$ (bottom). The progressively more red colors indicate a progressively larger number of points in a given interval of values of $B$ and the power of the above quantities, in logarithmic units.  Only points with plasma-$\beta$ below 10 and with magnetic field inclination $\theta<40\degr$ at heights between 1.1 and 1.3 Mm are selected for the analysis.}\label{fig:hifreq-b}
\end{figure*}

Figure \ref{fig:hifreq-b} confirms our visual impression about the correlations between the $f$-quantities and the magnetic field strength from Fig. \ref{fig:maps-hifreq-ambihall}. We applied a selection criterium on the points analyzed in this figure, selecting only those points at heights 1.1--1.3 Mm with $\beta<10$ and not strongly inclined fields with $\theta<40\degr$ to study potentially the regions where the Hall-induced transformation may be acting according to \citet{Cally+Khomenko2015, Gonzalez-Morales+etal2019}. We did not apply a more strict limit of $\beta<1$ because the amount of points satisfying both $\beta$ and $\theta$ criterium is not sufficiently large to make good statistics (notice that $B$ and $\beta$ are related, so narrowing the range of $\beta$ too much prevents from studying the dependence on $B$). 

We observe that for all the simulations the power of $\mathrm{f}_{\rm alf}$ increases with $B$ (top row). Therefore, whatever is the generation mechanism of these waves in our simulations, it seems to be similar for all three cases. Once generated, the amplitudes of $\mathrm{f}_{\rm alf}$  are larger for larger fields. This dependence extends to slightly higher field strengths in the \ambihall\ case, and the stronger fields allow one to have stronger waves amplitudes.  These strong waves are ``eaten out'' by the ambipolar diffusion in the \ambi\ case and are not renewed. The Hall effect apparently allows to renew them, contributing to the energy transport to the upper layers. A similar dependence is also observed for the Poynting flux, $\mathbf{S}_{\rm EM,z}^{\rm ideal}$,  and the magnetic field (bottom row). The flux is stronger when the field is stronger. Therefore, it can be concluded that stronger and more vertical fields in our simulations facilitate magnetic energy transport to the upper layers by means of Alfv\'en waves. 
 
The middle row of Fig. \ref{fig:hifreq-b}  shows rather an opposite dependence on the field for $\mathrm{f}_{\rm fast}$. The amplitudes of the fast waves either do not depend on the field for lower and vertical fields, or slightly decrease with the field for the stronger fields. This behavior is similar for all three cases. Compressible waves are more abundant in the lower field areas, for the same vertical inclination. This can be an indirect consequence of the correlation between the field strength and inclination, and their influence on the mode transformation.

\begin{figure*}
\centering
\includegraphics[keepaspectratio,width=16cm]{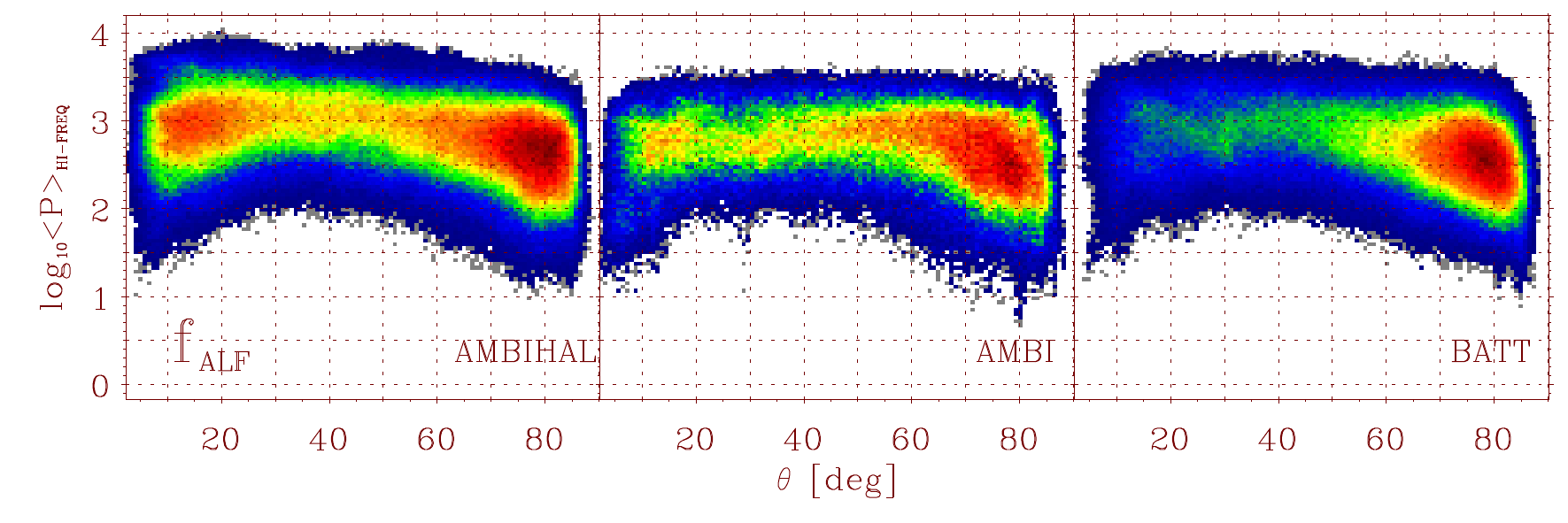}
\includegraphics[keepaspectratio,width=16cm]{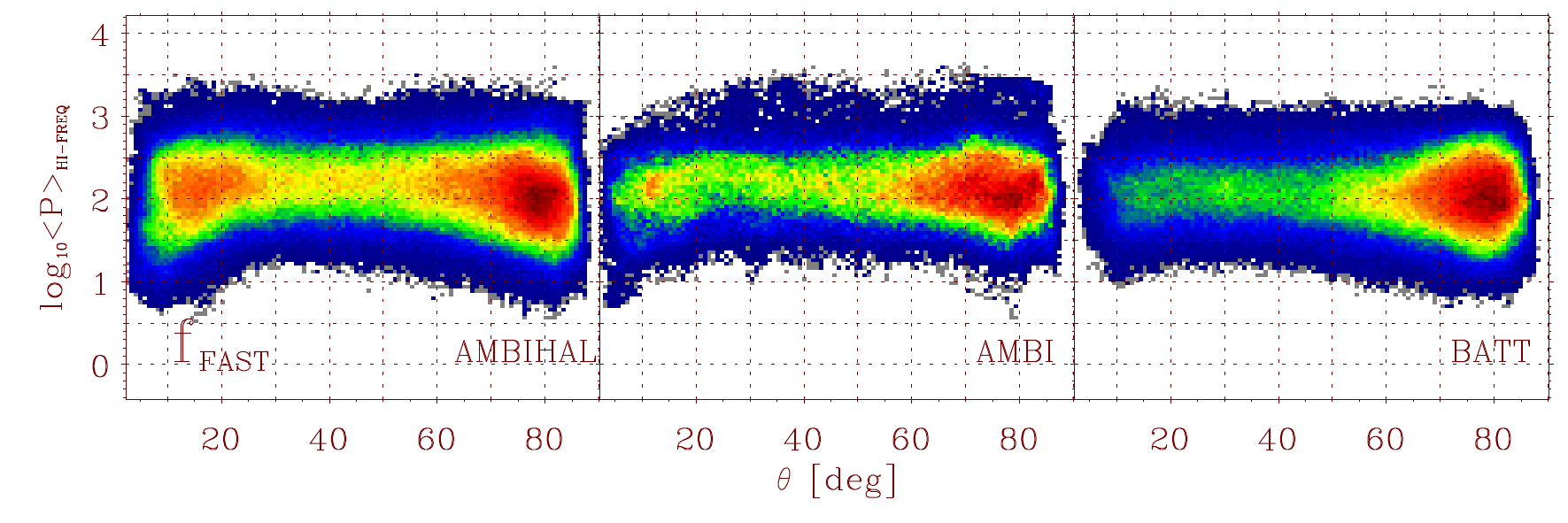}
\includegraphics[keepaspectratio,width=16cm]{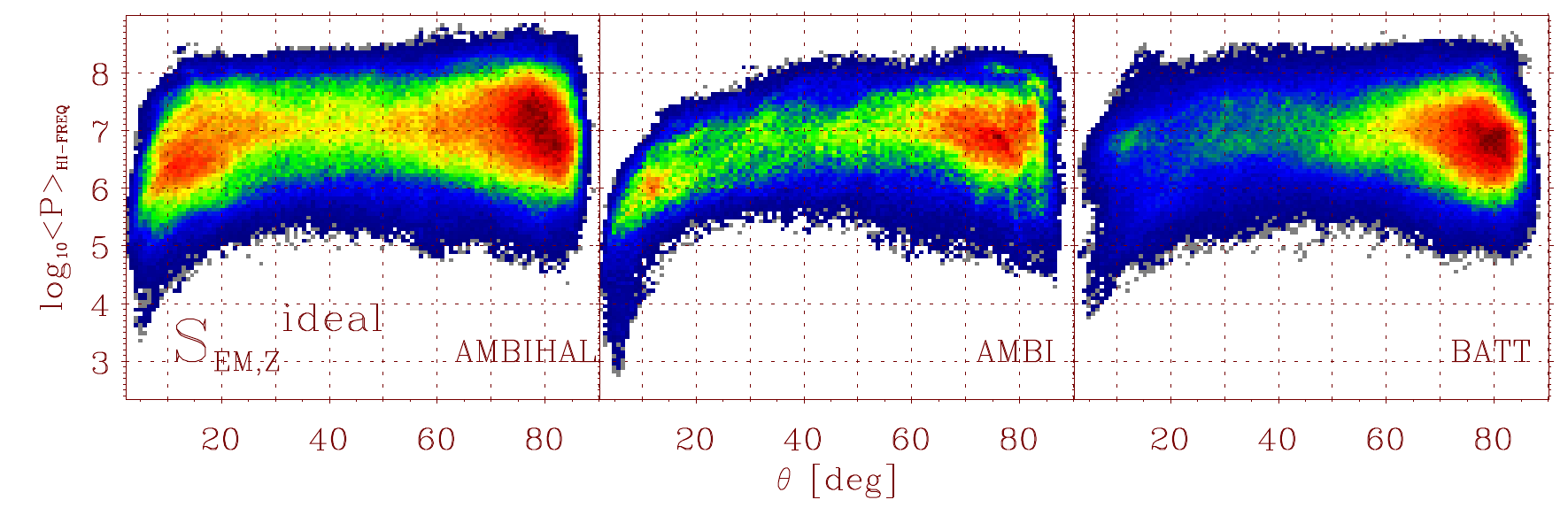}
\caption{\footnotesize  Two-dimensional histograms showing the dependence on magnetic field inclination, $\theta$, of the power of $\mathrm{f}_{\rm alf}$ (top), $\mathrm{f}_{\rm fast}$ (middle), and $\mathbf{S}_{\rm EM,z}^{\rm ideal}$ (bottom). The progressively more red colors indicate progressively a larger number of points in a given interval of values of $\theta$ and the power of the above quantities, in logarithmic units.  Only points with plasma-$\beta$ below 1 at heights between 1.1 and 1.3 Mm are selected for the analysis.} \label{fig:hifreq-theta}
\end{figure*}

Figure \ref{fig:hifreq-theta} shows the dependence between the power of oscillations and the inclination of the magnetic field. It is computed for those locations where plasma $\beta$ is below 1,  at heights between 1.1--1.3 Mm. This selection is made to make sure that compressible an incompressible perturbations are different enough and represent fast and Alfv\'enic waves, correspondingly, to as larger extent as possible. 

The distribution of $\mathrm{f}_{\rm alf}$ power (top row) has a bimodal shape for the \ambihall\ simulation, unlike the other two cases. There is a larger number of occurrences of points with vertical field in the \ambihall\ case, and these points also have slightly larger power of  $\mathrm{f}_{\rm alf}$. The latter may be an indication of the Hall-induced transformation because it works more efficiently for the vertical fields. In the \ambihall\ model we have, on the one hand, more vertical fields and, on the other hand, that the Hall parameter is large enough to allow for this transformation. The \ambi\ and \batt\ simulations show no dependence between the power of $\mathrm{f}_{\rm alf}$ (or $\mathrm{f}_{\rm fast}$) and the inclination for fields inclined below 50--60$\degr$, see upper and middle panels of Fig. \ref{fig:hifreq-theta} for these cases). 
The frequency of appearance of powerful oscillations in $f_{\rm alf}$ is significantly larger for large field inclinations. This seems to be in agreement with the geometrical mode transformation theory \citep{Cally+Goossens2008, Cally+Khomenko2012}.  Ambipolar diffusion can, in principle also affect waves by degrading the fast wave flux entering the transformation zone \citep{Cally+Khomenko2019, Cally+Khomenko2018, Raboonik+Cally2019}. This needs to be further investigated. 

The power of $\mathrm{f}_{\rm fast}$ (middle) is independent of the magnetic field inclination, but these compressible perturbations are, again, more probable to be found in vertical fields in the \ambihall\ model. This result can be explained by the fact that, for vertical fields, the Alfv\'en and fast waves are coupled and hardly distinguishable between each other. For the fluxes, $\mathbf{S}_{\rm EM,z}^{\rm ideal}$, the situation is similar. There is a large vertical flux for vertical fields in the \ambihall\ model, not present in the other two cases. For the \ambi\ and \batt\ models there are more instances of large $\mathbf{S}_{\rm EM,z}^{\rm ideal}$ for larger field inclination, probably as a consequence of the geometrical mode transformation.

\begin{figure*}
\centering
\includegraphics[keepaspectratio,width=16cm]{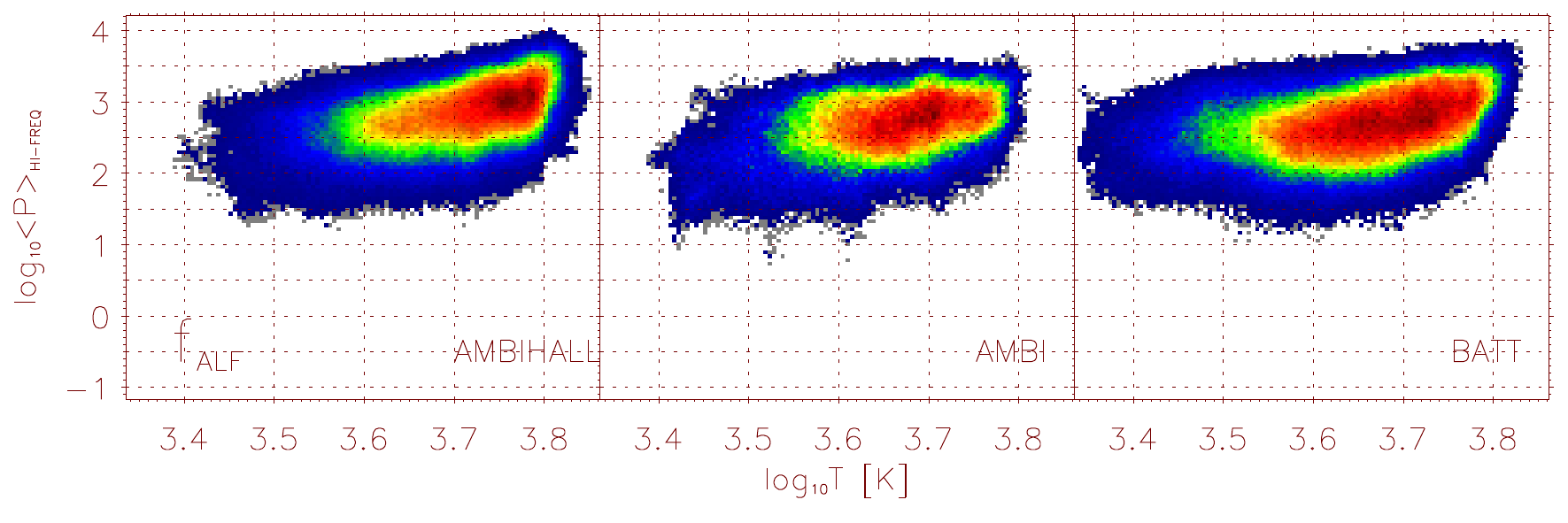}
\caption{\footnotesize Two-dimensional histograms showing the dependence on temperature of the power of $\mathrm{f}_{\rm alf}$. The progressively more red colors indicate a progressively larger number of points in a given interval of values of $T$ and the power of the above quantities, in logarithmic units.  Only points with plasma-$\beta$ below 1 at heights between 1.1 and 1.3 Mm are selected for the analysis. }\label{fig:hifreq-t}
\end{figure*}

\begin{figure*}
\centering
\includegraphics[keepaspectratio,width=16cm]{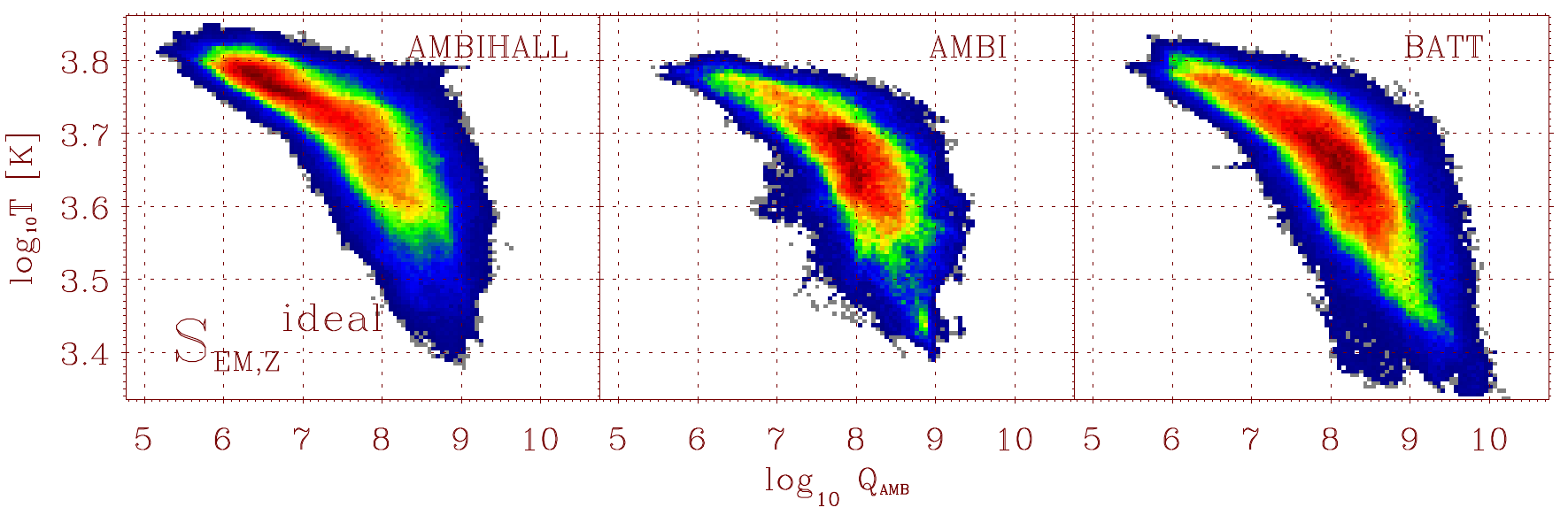}
\caption{\footnotesize Two-dimensional histograms showing the relation between the temperature and the ambipolar heating rate, $Q_{\rm AMB}$. The progressively more red colors indicate a progressively larger number of points in a given interval of values of $T$ and $Q_{\rm AMB}$, in logarithmic units.  Only points with plasma $\beta$ below 1 at heights between 1.1 and 1.3 Mm are selected for the analysis. }\label{fig:hifreq-qamb}
\end{figure*}

\begin{figure*}
\centering
\includegraphics[keepaspectratio,width=16cm]{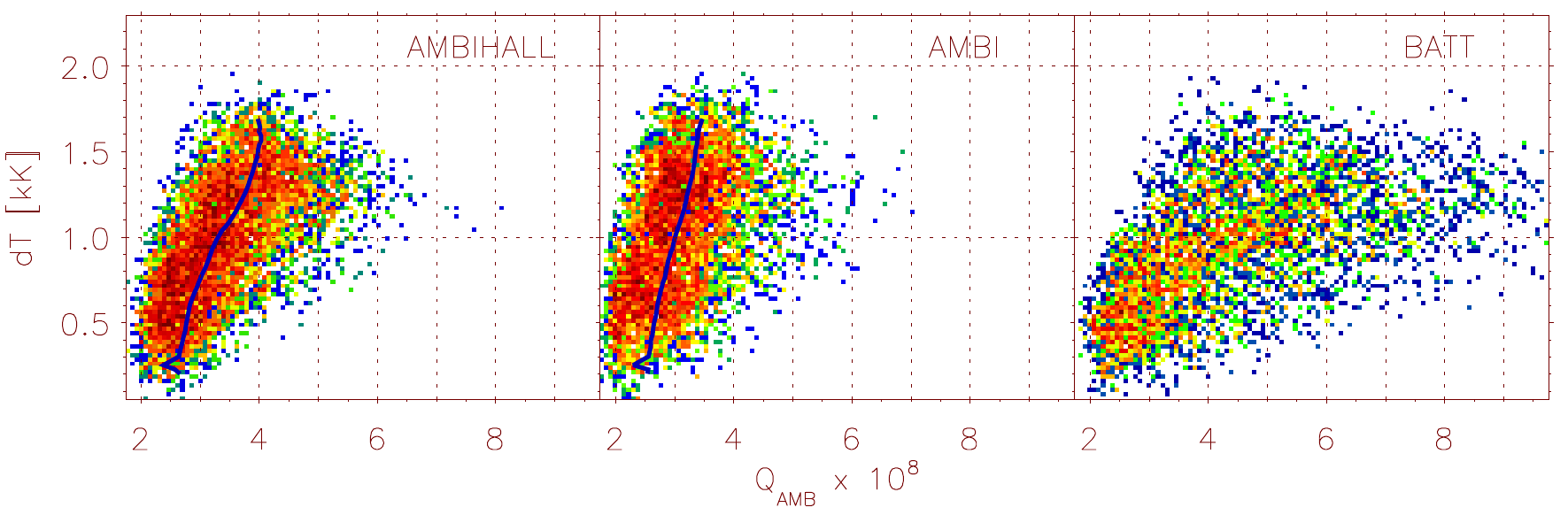}
\caption{\footnotesize Two-dimensional histograms showing the relation between the mean temperature difference between the two snapshots separated 100 sec in time, and the mean ambipolar heating rate, $Q_{\rm AMB}$ at heights between 0.56 Mm and 1.1 Mm and at locations where $Q_{\rm AMB}$ exceeds a certain value. The progressively more red colors indicate a progressively larger number of points in a given interval of values of $dT$ and $Q_{\rm AMB}$.   }\label{fig:dt-qamb}
\end{figure*}


Figure \ref{fig:hifreq-t} represents the dependence of $\mathrm{f}_{\rm alf}$ on temperature and is done for locations with $\beta$ below 1. It can be noted that there is a lower frequency of occurrence of points with low temperature in the \ambi\ and \ambihall\ runs. This is seen in the diagrams as the extension toward lower values of temperature in the \batt\ case (right).  Simultaneously,  for these locations with low plasma $\beta$, there are more occasions with higher temperature associated with the high amplitude of Alfv\'enic $\mathrm{f}_{\rm alf}$ perturbations  in the \ambihall\ case (left). This does not happen in the \ambi\ case because the highest amplitude waves are probably dissipated by the ambipolar effect, and are not reproduced again, unlike in the \ambihall\ case. The heating is more efficient in \ambihall\ because it is done via destruction of waves with higher amplitudes, while in the \ambi\ case this is done by the destruction of waves with lower amplitudes, since there is no supply of high-amplitude waves.

The location of points with most probable values of $T$ and $\mathrm{f}_{\rm alf}$ power (dark red color) is also different between the simulations. The most probable values are concentrated around the highest values of both $T$ and $\mathrm{f}_{\rm alf}$ power for the \ambihall\ case, indicating that locations with highest temperatures and wave power are cospatial and more frequent in this case, unlike the other cases.

The temperature and the power of $\mathrm{f}_{\rm fast}$ (as well as the power of $\mathbf{S}_{\rm EM,z}^{\rm ideal}$) are independent (not shown in Fig. \ref{fig:hifreq-t}). There are more locations with higher temperature corresponding to large $\mathbf{S}_{\rm EM,z}^{\rm ideal}$ in the \ambihall\ case. In general, it should be noted that it is hard to establish the dependence between the Poynting flux and the temperature. In principle, larger fluxes are not expected to generate larger in situ temperatures. The dependence between the flux and the temperature may rather reflect the correlations between the magnetic field and the granulation pattern.

The latter point is illustrated in Figure \ref{fig:hifreq-qamb}. This figure shows the dependence between the ambipolar heating rate, $Q_{\rm AMB}=\eta_AJ_\perp^2$, and the temperature.  One can observe that larger temperatures correspond to lower ambipolar heating. Such dependence is because the ambipolar heating happens in the low temperature areas, where the amount of neutrals is larger, and the magnetic field is higher (as e.g., intergranular lanes or rarefactions corresponding to chromospheric shocks). From the result above we have seen that  places with stronger and vertical fields harbor more powerful waves. The dissipation of such waves should produce larger heating and temperatures, but this is not reflected in Figure \ref{fig:hifreq-qamb}. 

In order to mitigate this unsatisfactory result, we used the fact that ambipolar heating has a characteristic time on the order of $10^1-10^2$ sec \citep{Khomenko+Collados2012, Nobrega-Siverio+etal2020}. Using the whole time span of the simulations, we have picked up the spatial locations where $Q_{\rm AMB}$ was above a certain threshold at heights between 0.56 and 1.1 Mm. At these locations we averaged the value of the $Q_{\rm AMB}$, and the value of the temperature difference between two snapshots separated 100 sec in time. The results are given at Figure \ref{fig:dt-qamb}. This figure clearly shows that when the ambipolar diffusion is included (left and middle panels), there is a positive correlation between $Q_{\rm AMB}$ and the temperature increase at these location 100 sec after. The time interval of 100 sec gives the best correlation, though similar results are obtained for the time intervals in the range 50-200 sec. The \batt\ simulation shows clearly more dispersion and no sign of temperature increase for large $Q_{\rm AMB}$.

\section{Discussion and conclusions}


We have analyzed the first 3D simulations of solar dynamo extending to chromospheric heights that include the action of three most important nonideal effects: ambipolar diffusion, the Hall effect, and Biermann battery effect. The simulations last for several solar hours in the stationary regime, which has allowed us to reach a statistically valid state, and make sure that the action of nonideal effects is fully developed. We have improved our previous work (Paper I) by including the Hall effect into consideration. We have also increased the statistics of our study by considering a two times longer series (2 hours) with twice higher cadence of snapshot saving (10 seconds). The analysis has allowed us to study perturbations in a wide range of frequencies, up to 50 mHz. Our main conclusions can be summarized in the following way:

\begin{itemize}
\item[(i)] Hall and ambipolar effects influence the average temperature structure of the simulations. Simulations with both, Hall and ambipolar effects, result in larger temperatures in the low chromosphere (0.8--1 Mm), while the simulations where solely the ambipolar effect is acting produce a temperature increase at higher heights. There is a positive correlation between the magnitude of the ambipolar heating term, $Q_{\rm AMB}$, and the temperature increase $\sim10^2$ sec later at the same location in simulations where the ambipolar effect is acting.

\item[(ii)]  The magnetic field is lower and more horizontal in the chromosphere in the simulations with the ambipolar effect only, being consistent with the picture of dissipation of magnetic energy and its conversion into the thermal energy. The Hall effect causes the opposite effect: its continuous action in our model results in stronger, more vertical, and long-living magnetic field concentrations. The field is on average stronger in the chromosphere in the simulations with the Hall effect. 

\item[(iii)] The width of the current sheets is smaller in the simulations where only the ambipolar diffusion is acting, but the addition of the Hall effects results in an increase.

\item[(iv)]  Both Hall and ambipolar effects influence the power of compressible waves in the chromospheric layers of  our simulations. There is a 30--40\% increase of power for compressible perturbations above 0.5 Mm in the \ambi\ simulations, relative to the \batt\ simulations. As a result, there is a large abundance of fast magneto-acoustic shocks at chromospheric levels in the \ambi\ case. Unlike that, there is a 30--40\% less power of compressible perturbations above 0.5 Mm in the \ambihall\ simulations relative to \ambi\ simulations. Both effects increase with frequency. 

\item[(v)]  The power of incompressible (Alfv\'en) waves is decreased by 20--30\% in the \ambi\ simulations relative to the \batt\ simulations, that is, the ambipolar effect allows to dissipate these incompressible perturbations. However, the \ambihall\ simulation with the Hall effect included produces about twice larger power of these incompressible waves in the chromospheric layers. The waves dissipated by the ambipolar effect are continuously replaced by the action of the Hall effect. The effects are more pronounced at lower frequencies. 

\item[(vi)]  The ambipolar effect is responsible for the Poynting flux absorption through the dissipation of incompressible (Alfv\'en) waves. Some 30--40\% of the Poynting flux reaching the chromosphere is absorbed in the \ambi\ model relative to the \ambihall\ model. The presence of the Hall effect results in about twice larger Poynting flux in the chromosphere in the \ambihall\ case relative to the \ambi\ case. 

\item[(vii)]  The power of Alfv\'en waves is larger for stronger and more vertical fields, while the opposite is observed for the compressible perturbations associated with the fast MHD waves. The presence of stronger, more vertical and long-living magnetic field concentrations in the \ambihall\ model helps the Poynting flux reaching the upper layers. Therefore, the Hall effect indirectly helps Alfv\'en waves reaching the chromosphere by creating such structures. 

\item[(viii)]  There are indications that more Alfv\'en waves are produced at locations with vertical field in the \ambihall\ model which can be attributed to the Hall-induced mode transformation. Nevertheless, we were unable to find more direct confirmations of the presence of the Hall-induced transformation due to the natural correlations between the Hall parameter and the temperature and magnetic structure of granulation. 

\item[(ix)]  There are evidences for larger-amplitude Alfv\'en waves producing more instances with larger temperatures in the simulations with the Hall effect. This result confirms the role of Alfv\'en waves in heating chromospheric flux tubes. 

\end{itemize}

The Hall effect is frequently considered not important for simulations of large-scale flows in a plasma, as the ones considered here. This is so because the Hall effect is supposed to act at small temporal and spatial scales. The Hall effect in a fully ionized plasma arises because the ions are relatively less mobile than electrons. This produces a drift between ions and electrons in the direction of the Lorentz force. In a fully ionized plasma, the Hall effect acts at frequencies similar to the ion-cyclotron frequency. The latter is on the order of MHz for the typical magnetic fields of the solar atmosphere \citep{Khomenko+etal2014}.  However, in partially ionized plasmas the action of the Hall effect extends to lower frequencies, because the Hall coefficient is scaled with the ionization fraction, which can be very low, see Eq. (\ref{eq:epsilon1}) \citep{Pandey+Wardle2008, Pandey2008, Cheung+Cameron2012}. In a partially ionized plasma, where the collisions dominate, as in the solar photosphere and the low chromosphere, the ions become dragged (or slowed down) by neutrals which do not directly feel the Lorentz force and this creates Hall currents of much lower frequencies. The Hall effect in the partially ionized solar plasma may become important at time scales of the order to $10^{-2}$--$10^{-3}$ seconds, which are already time scales resolved in simulations. 

There are two works where the Hall effect has been included in magneto-convection simulations, see \citet{MartinezSykora+etal2012} and \citet{Cheung+Cameron2012}. In both cases, only 2D simulations have been considered, with relatively short time series. In the simulations of the umbra magneto-convection, \citet{Cheung+Cameron2012} describe the generation of a small, out of the 2D plane, component of the velocity and magnetic field. Once generated, this magnetic field gets advected by convective rolls and is dragged to the convection zone. The field generated through this effect was on the order of 5 G, compared to the kG field in the umbra, and the associated velocities were on the order of 100 m/s \citet{Cheung+Cameron2012}. While these values, by themselves, do not seem large, this effect was not present when the Hall term was not included. The Hall effect is an intrinsically 3D effect, because it creates velocities in the direction perpendicular to both, the magnetic field, and the (arbitrary) initial perturbation,
$\mathbf{v}_{\rm HALL}\sim \mathbf{\nabla}\times\mathbf{B}$. 
Therefore, its full action can only be adequately considered in 3D experiments, as the ones reported here. 

The continuous action of the Hall effect in a 3D situation has the capability to lead to field amplification. The possibility of strong flux tubes formation thanks to the Hall effect was suggested by \citet{Khodachenko+Zaitsev2002}. This mechanism works at locations with a converging flow of partially ionized plasma due to the balance between the Lorentz force and the ion-neutral collisional force. The intensification is likely to begin in the upper layers and to propagate downwards. The action of \citet{Khodachenko+Zaitsev2002} mechanism in our simulations needs to be investigated further. In a subsequent paper, \citet{Khomenko2020} traced an individual long-living feature in the \ambihall\ simulation. The results show that there is an enhancement of the $\eta_H$ coefficient right around the feature coincident with local intensification of the magnetic field. However, a more statistical investigation is needed to claim the relation between the formation of these features and the Hall effect.

The effects of the Hall-induced mode transformation in our simulations are less obvious. We clearly see that the power of compressible and incompressible waves depends on magnetic field strength and inclination. The incompressible waves are more powerful in vertical magnetic structures that live longer in the \ambihall\ simulation. Nevertheless, the magnetic field is too weak in the simulations to have an ample zone with low plasma-$\beta$ in the chromosphere. In order to produce Alfv\'en waves through the geometric mode transformation mechanism \citep{Cally+Goossens2008, Khomenko+Cally2012}, fast mode waves should be produced first at the $\beta=1$ layer and then these should be able to refract. It is not likely that a large amount of Alfv\'en waves is produced this way, because the fast mode refraction height is probably mostly outside our domain. The Hall-induced transformation is probably taking place because the values of the Hall parameter $\epsilon$ are large at the upper part of the domain. However, the correlations between the power of incompressible Alfv\'en waves and the Hall parameter are the opposite to what is expected for this mechanism, and rather reflect correlations that are natural for the granulation structure. Therefore, we consider as the most plausible scenario that most of these incompressible perturbations have been produced already at photospheric layers with large plasma-$\beta$ by granulation, and their larger presence in the upper layers of the \ambihall\ model is the consequence of the large fraction of stronger and vertical magnetic structures there. 

One of the most important conclusions of our work is about the importance of the Hall effect for the chromospheric energy balance. The Hall effect by itself does not produce dissipation and can only redistribute magnetic energy.  However, we have seen that its action allows to supply more powerful incompressible Alfv\'en waves to the chromosphere. These waves are dissipated by the ambipolar mechanism and reproduced back again thanks to the Hall effect. We have obtained unequivocal evidences that high-frequency Alfv\'en waves heat magnetic flux tubes in the chromosphere. There is a larger occurrence of hotter-temperature instances in the chromosphere, cospatial with large power of incompressible perturbations, in the \ambihall\ model, and there is evidence that the temperature is increased at locations with large $Q_{\rm AMB}$ after some characteristic time around 100 sec.


There are few drawbacks in our modeling that need to be improved in the future to make ultimate conclusions about the role of partial ionization effects on the chromospheric energy balance. The most important issue is our treatment of radiative losses and ionization balance. It has been repeatedly shown \citep{Leenaarts2006, Leenaarts2007, Wedemeyer2011}, that characteristic ionization and recombination time scales of hydrogen, helium, and other elements in the chromosphere are out of balance and comparable to hydrodynamical time scales. This results in overionization of the chromosphere in respect to the case of instantaneous equilibrium assumed in our work. In their 2D experiments of flux emergence, \citet{Nobrega-Siverio+etal2020} have shown that taking these  nonequilibrium ionization effects into account may result in an insufficient action of the ambipolar diffusion for heating cold rising bubbles produced during the emergence. Nevertheless, these authors found that an efficient ambipolar heating is still produced in shocks. Taking nonequilibrium ionization, and the ambipolar and the Hall effects together in 3D modeling is extremely numerically costly. This problem becomes especially severe because, as we have seen here, long time series must be analyzed to make sure that a stationary situation is fully achieved. This effort must be undertaken in future studies. 

The physical situation considered in this work (small-scale dynamo) is different from the models of flux emergence, as in \citet{Nobrega-Siverio+etal2020}. In small-scale dynamo simulations, the fields are significantly weaker and structured on much smaller scales, and cool ascending bubbles are not produced. The presence of small-scale structures create currents and allows to dissipate energy through ambipolar mechanism efficiently enough already at the upper photosphere \citep{Khomenko+etal2018}, which should be less affected by the issues of time-dependent ionization.  The action of Hall and ambipolar mechanisms may differ in such different physical situations. The evaluation of the role of these mechanisms for the chromosphere must be done by considering simulations with different magnetization level, and with nonzero vertical magnetic flux, as for example models of plage and sunspot regions. The dependence of the ambipolar heating rates with the magnetic flux level may have an impact onto solar irradiance variations \citep[see the discussion in][]{Rempel2020}, and its study would be a recommendable next step for future studies.

The numerical resolution of simulations must also be improved in the future. The importance of all the nonideal effects increases with decreasing spatial and temporal scales. Therefore, simulations at different resolutions must be compared in the future. \\












{\bf Acknowledgements.} 
This work was supported by the Spanish Ministry of Science through the project PGC2018-095832-B-I00. It contributes to the deliverable identified in FP7 European Research Council grant agreement ERC-2017-CoG771310-PI2FA for the project ``Partial Ionization: Two-fluid Approach''. The authors thankfully acknowledge the technical expertise and assistance provided by the Spanish Supercomputing Network (Red Espa\~nola de Supercomputaci\'on), as well as the computer resources used: LaPalma Supercomputer, located at the Instituto de Astrof\'isica de Canarias, and MareNostrum based in Barcelona/Spain.


\providecommand{\noopsort}[1]{}\providecommand{\singleletter}[1]{#1}%

\end{document}